\documentclass[sigplan,10pt]{acmart}
\acmConference[PL'18]{ACM SIGPLAN Conference on Programming Languages}{January 01--03, 2018}{New York, NY, USA}
\acmYear{2018}
\acmISBN{} 
\acmDOI{} 
\startPage{1}

\setcopyright{none}

\usepackage{booktabs}   
\usepackage{subcaption} 

\usepackage{booktabs}   
\usepackage{subcaption} 
\usepackage{amsmath,comment}
\usepackage{color,soul}

\definecolor{bg}{rgb}{0.95,0.95,0.95}
\definecolor{darkgreen}{RGB}{0,124,0}
\definecolor{lightgray}{RGB}{211,211,211}

\definecolor{highlightcolor}{rgb}{0.85,0.85,0.85}
\sethlcolor{highlightcolor}

\newcommand{\myhl}[1]{\mbox{\hl{$#1$}}}

\usepackage{mathtools}
\usepackage{mathrsfs}
\usepackage{blindtext}
\usepackage[frozencache]{minted}
\RecustomVerbatimEnvironment{Verbatim}{BVerbatim}{} 

\usepackage{eqnarray}

\usepackage{amssymb}
\usepackage{mathtools}

\usepackage{ulem}
\usepackage{array}

\usepackage{colortbl}
\usepackage{multirow}

\DeclarePairedDelimiter\JVMState{\lBrack}{\rBrack}
\newcommand{\tool}{\textsc{DeJITLeak}}
\newcommand{\toollight}{\textsc{DeJITLeak$_{\tt light}$}}
\newcommand{\JITFlow}{\textsc{JITFlow}}

\newcommand{\Methods}{\mathbf{M}}
\newcommand{\pc}{\textsf{pc}}
\newcommand{\os}{\textsf{os}}
\newcommand{\cs}{\textsf{cs}}
\newcommand{\argv}{\textsf{argv}}
\newcommand{\push}{\textsf{push}}
\newcommand{\invoke}{\textsf{invoke}}
\newcommand{\pop}{\textsf{pop}}
\newcommand{\binop}{\textsf{binop}}
\newcommand{\swap}{\textsf{swap}}
\newcommand{\load}{\textsf{load}}
\newcommand{\store}{\textsf{store}}
\newcommand{\heap}{\textsf{h}}
\newcommand{\cheap}{\textsf{ch}}
\newcommand{\mm}{\textsf{mm}}

\newcommand{\base}{\textsf{base\_version}}
\newcommand{\putt}{\textsf{put}}
\newcommand{\get}{\textsf{get}}
\newcommand{\sub}{\textsf{sub}}
\newcommand{\ifeq}{\textsf{ifeq}}
\newcommand{\ifneq}{\textsf{ifneq}}
\newcommand{\goto}{\textsf{goto}}
\newcommand{\deopt}{\textsf{deopt}}
\newcommand{\return}{\textsf{return}}

\newcommand{\statejvm}{\mathbf{States}}
\newcommand{\Conf}{\mathbf{Conf}}
\newcommand{\LVar}{\mathbf{LVar}}
\newcommand{\GVar}{\mathbf{GVar}}
\newcommand{\Val}{\mathbf{Val}}
\newcommand{\newjvm}{\textsf{new}}
\newcommand{\getfield}{\textsf{getfield}}
\newcommand{\putfield}{\textsf{putfield}}

\newcommand{\dir}{\mathbf{d}}
\newcommand{\Dir}{\mathbf{D}}
\newcommand{\md}{\texttt{md}}

\newcommand{\prot}{\texttt{prot}}

\newcommand{\sig}{\textsf{sig}}
\newcommand{\pt}{\textsf{pt}}  
\renewcommand{\st}{\textsf{st}} 
\newcommand{\heapt}{\textsf{ht}}  
\newcommand{\lt}{\textsf{lt}}  
\newcommand{\nextt}{\textsf{nxt}}
\newcommand{\junc}{\textsf{junc}}

\newcommand{\seclev}{\mathbb{L}}
\newcommand{\high}{\textcolor{red}{\textsf{H}}}
\newcommand{\low}{\textcolor{blue}{\textsf{L}}}

\newcommand{\transform}{\textsf{T}}
\newcommand{\prof}{\textsf{pf}}
\newcommand{\cost}{\textsf{cf}}
\newcommand{\se}{\textsf{se}}
\newcommand{\proj}{\textsf{proj}}

\newcommand{\region}{\textsf{region}}
\newcommand{\maxbp}{\textsf{maxBP}}
\newcommand{\ifb}{\textsf{if-b}}
\newcommand{\elseb}{\textsf{else-b}}

\newcommand{\qq}[1]{\textcolor{purple}{#1}}

\begin{document}

\title{Preventing Timing Side-Channels via Security-Aware Just-In-Time Compilation}         


\author{Qi Qin}
\affiliation{
  \institution{ShanghaiTech University}            
  \city{Shanghai}
  \country{China}                    
}

\author{JulianAndres JiYang}
\affiliation{
  \institution{ShanghaiTech University}            
  \city{Shanghai}
  \country{China}                    
}

 \author{Fu Song}
\affiliation{
  \institution{ShanghaiTech University}            
  \city{Shanghai}
  \country{China}                    
}

 \author{Taolue Chen}
\affiliation{
  \institution{Birkbeck, University of London}            
  \city{London}
  \country{United Kingdom}                    
}
 \author{Xinyu Xing}
\affiliation{
  \institution{Northwestern University}            
  \city{Evanston}
  \country{USA}                    
}
\begin{abstract}
Recent work has shown that Just-In-Time (JIT) compilation can introduce timing side-channels to constant-time programs, which would otherwise be a principled and effective means to counter timing attacks.
In this paper, we propose a novel approach to eliminate JIT-induced leaks from these programs. Specifically, we present an operational semantics and a formal definition of constant-time programs under JIT compilation,
laying the foundation for reasoning about programs with JIT compilation.
We then propose to eliminate JIT-induced leaks via a fine-grained JIT compilation for which we provide an automated approach
to generate policies and a novel type system to show its soundness.
We develop a tool {\tool} for Java based on our approach and implement
the fine-grained JIT compilation in HotSpot. Experimental results show that {\tool} can effectively and efficiently eliminate JIT-induced leaks on three datasets used in side-channel detection.
\end{abstract}



\maketitle

\section{Introduction} \label{sect:intro}

Timing side-channel attacks allow an adversary to infer secret information by measuring the execution time of an implementation,
thus pose a serious threat to secure systems~\cite{Kocher96}.
One notorious example is Lucky 13 attack that can remotely recover plaintext from the CBC-mode encryption in TLS due to an unbalanced branch statement~\cite{AlFardanP13}.

Constant-time principle, which requires the execution time of an implementation being independent of secrets,
is an effective countermeasure to prevent such attacks.
However, writing constant-time programs is error-prone.
For instance, even though two protections against Lucky 13 were implemented in AWS's s2n library,
a variant of Lucky 13 can remotely and completely recover plaintext from the CBC-mode cipher suites in s2n~\cite{AlbrechtP16}.
Therefore, various approaches have been proposed for automatically verifying constant-time security of high-/intermediate-level programs, e.g., ct-verif~\cite{AlmeidaBBDE16} and CacheAudit~\cite{DoychevFKMR13} for C programs,
CT-Wasm~\cite{WattRPCS19} for WASM programs,
Blazer~\cite{AntonopoulosGHK17} and Themis \cite{ChenFD17} for Java programs,
and FaCT~\cite{CauligiSJBWRGBJ19} for eliminating leakages.

However, constant-time programs may be still vulnerable in practice if the runtime environment is not fully captured by constant-time models. For instance, static compilation from high-/intermediate-level programs to low-level counterparts
can destruct constant-time security~\cite{BartheGL18,BartheBGHLPT20,BartheBHP21};
constant-time executable programs are vulnerable in modern processors due to, e.g., speculative or out-of-order execution~\cite{KocherHFGGHHLM019,Lipp0G0HFHMKGYH18,CauligiDGTSRB20};
JIT compilation makes constant-time bytecode vulnerable~\cite{BrennanRB20,BrennanSB20},
called JIT-induced leaks hereafter. In this work, we focus on JIT-induced leaks.

JIT compilation has been used in numerous programming language engines
(e.g., PyPy for Python, LuaJIT for Lua, HotSpot for Java, and V8 for JavaScript) to improve performance.
However, JIT compilation can break the balance of conditional statements, e.g.,
some methods are JIT compiled or inlined in one branch but not in the other branch,
or one branch is speculatively optimized, making constant-time programs vulnerable at runtime as shown in~\cite{BrennanRB20,BrennanSB20}. Despite the serious risk of JIT compilation, there is no rigorous approach to eliminate
JIT-induced leaks otherwise completely disabling JIT compilation.

In this work, we aim to automatically and rigorously eliminate JIT-induced leaks.
Our contributions are both theoretical and practical. On the theoretical side,
we first lay the foundations for timing side-channel security under JIT compilation
by presenting a formal operational semantics and defining a notion of constant-time for a fragment of the JVM under JIT compilation.
We do not model concrete JIT compilation as done by~\cite{FluckigerSYGAV18, BarriereBFPV21}.
Instead, we leave them abstract in our model and model JIT compilation via compilation directives controlled
by the adversary. This allows  to consider very powerful attackers who have control over
JIT compilation. It also makes it possible to reason about bytecode running with JIT compilation
and uncover how code can leak secrets due to JIT compilation in a principled way.
We then propose to prevent JIT-induced leaks via a fine-grained JIT compilation
and present a type system for statically checking the effectiveness of
policies for fine-grained JIT compilation.

On the practical side, we propose \tool, an automatic technique to generate policies
that can be proven to completely eliminate JIT-induced leaks, while still benefiting from the performance gains of JIT compilation;
in addition, a lightweight variant of \tool, \toollight, can eliminate most of the leaks with a low overhead for more performance-conscious applications
and is still sound if methods invoked in both sides of each secret branching statement are the same. We implement {\tool} as a tool and fine-grained JIT compilation in HotSpot JVM from OpenJDK.
We conduct extensive experiments on three datasets used in recent side-channel detection: DifFuzz \cite{nilizadeh2019diffuzz}, Blazer \cite{AntonopoulosGHK17} and Themis \cite{ChenFD17}.
Experimental results show that our approach significantly outperforms the  strategies proposed in~\cite{brennan2020static}.
We report interesting case studies which shed light on directions for further research in this area.

In summary, our contributions are:
\begin{itemize}
  \item A formal treatment of JIT-induced leaks including an operational semantics and a constant-time notion under JIT compilation;
  \item A protection mechanism against JIT-induced leaks via a fine-grained JIT compilation and an efficient approach to generate policy for fine-grained JIT compilation with security guarantees;
  \item A practical tool that implements our approach and extensive experiments to demonstrate the efficacy of our approach.
\end{itemize}

\section{Overview} \label{sect:overview}
In this section, we first give a brief overview of the side-channel leaks induced by JIT compilation~\cite{BrennanRB20}.
We will exemplify these JIT-induced leaks using the HotSpot virtual machine (HotSpot for short) on OpenJDK 1.8.
We then give an overview of our approach to identify and eliminate the JIT-induced leaks automatically.

\begin{figure*}
  \centering
\begin{subfigure}[b]{0.24\textwidth}
\centering
\begin{minted}[fontsize=\scriptsize,bgcolor=bg]{java}
boolean pwdEq(char[] a,char[] b){
    boolean equal = true;
    boolean shmequal = true;
    for (int i = 0; i < 8; i++) {
        if (a[i] != b[i])
            equal = false;
        else shmequal = false;
    }
    return equal; }
\end{minted}
  \caption{The \textcolor{blue}{pwdEq} method}
  \label{fig:ToptiDemo-prog}
  \end{subfigure}
  \begin{subfigure}[b]{0.24\textwidth}
    \centering
    \includegraphics[width=\textwidth]{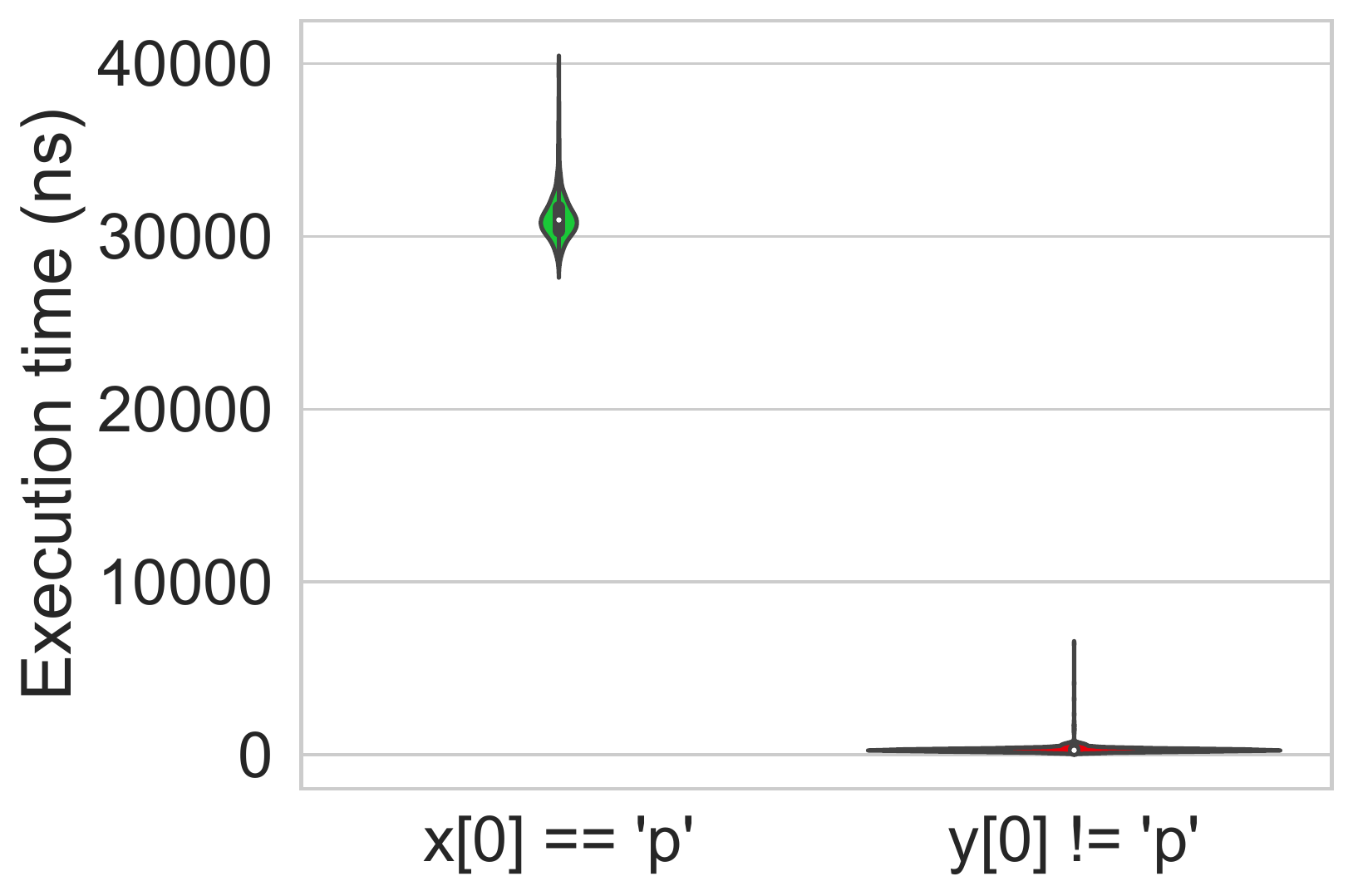}
    \caption{JIT enabled}
    \label{fig:ToptiDemo-JITenabled}
  \end{subfigure}
    \begin{subfigure}[b]{0.24\textwidth}
    \centering
    \includegraphics[width=\textwidth]{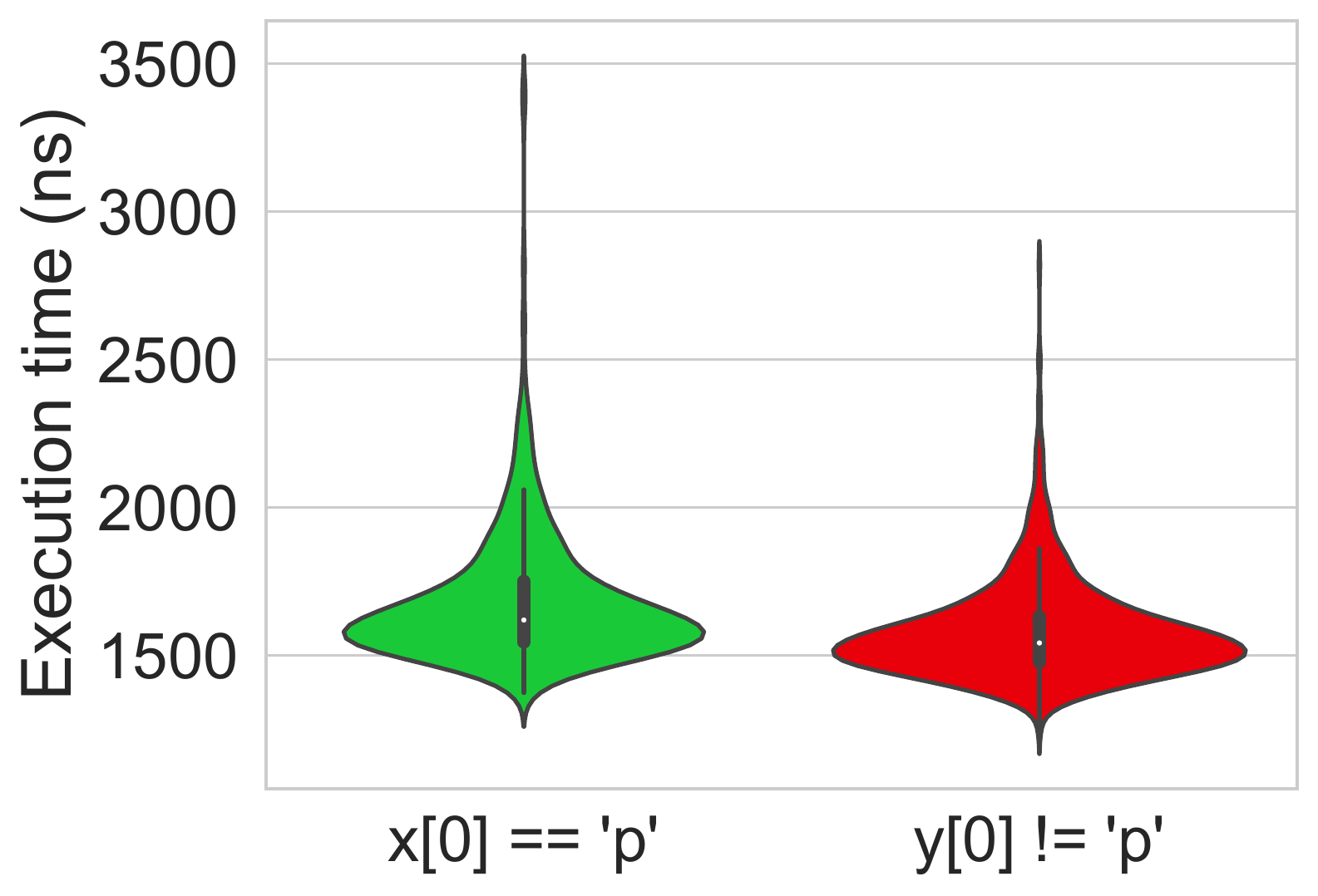}
    \caption{JIT disabled}
    \label{fig:ToptiDemo-JITdisabled}
  \end{subfigure}
    \begin{subfigure}[b]{0.24\textwidth}
    \centering
    \includegraphics[width=\textwidth]{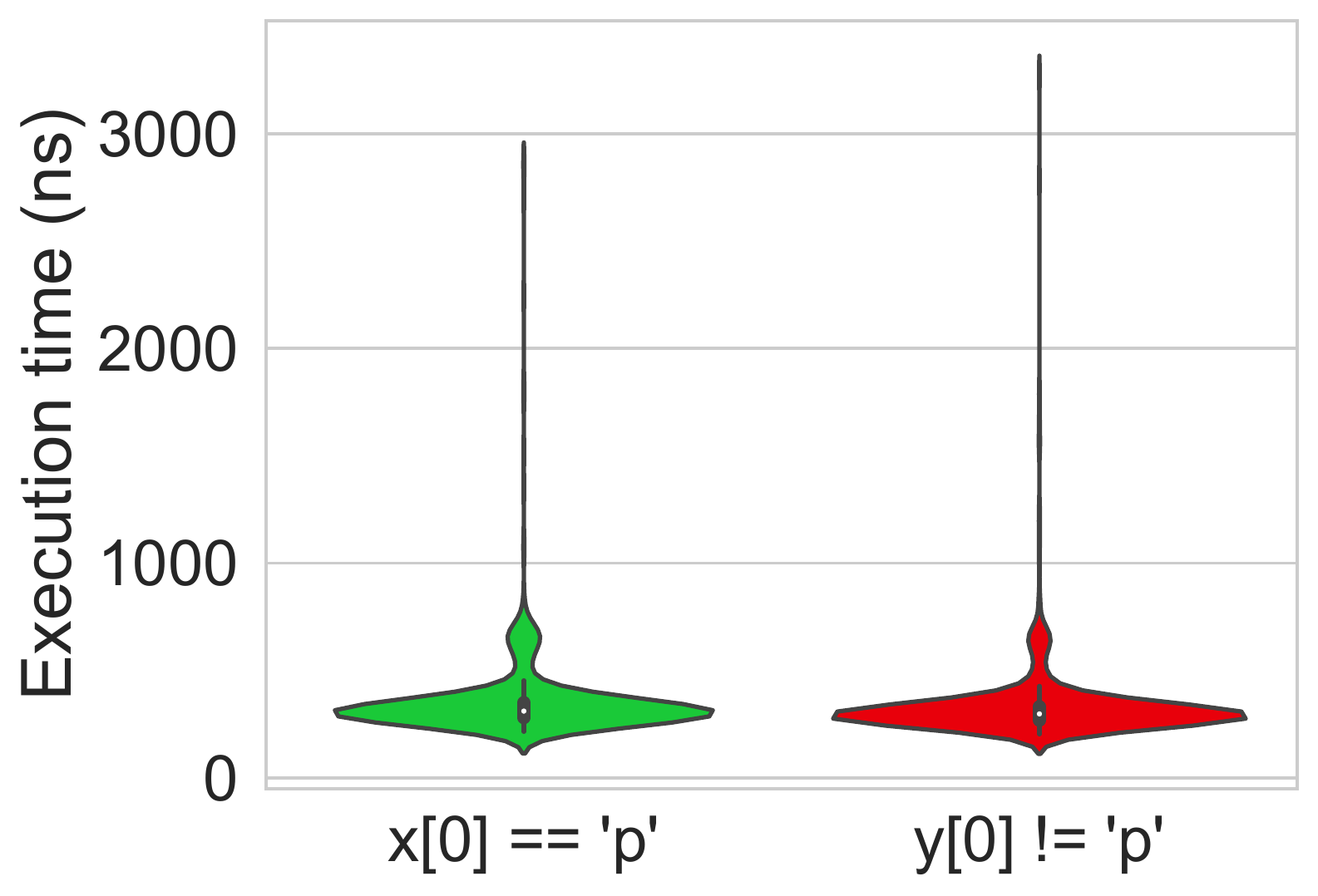}
    \caption{Mitigated}
    \label{fig:mitigationDemopwdEq}
  \end{subfigure}
  \vspace*{-2mm}
  \caption{The \textcolor{blue}{pwdEq} method and its execution time with JIT enabled and disabled under \textsc{Topti}}
  \centering \vspace*{-1mm}
  \label{fig:ToptiDemo}
\end{figure*}


\subsection{JIT-Induced Leaks}\label{sec:overviewjitleaks}
JIT-induced leaks could be caused at least by the following three JIT compilation techniques, i.e.,
(1) optimistic compilation, (2) branch prediction, and
(3) method compilation.
%

\smallskip\noindent{\bf Optimistic compilation (\textsc{Topti}).}
Optimistic compilation is a type of speculation optimizations~\cite{BarriereBFPV21}.
During the JIT compilation of a method, the compiler speculates on the most likely executed branches by pruning rarely executed branches.
Therefore, it reduces the amount of time required to compile methods at runtime
and space to store the native code.
However, there might be a subsequent execution where the speculation fails and the execution must fall back to bytecode in the interpreted mode. To handle this issue,
a deoptimization point (known as an uncommon trap in HotSpot) is added to the native code and, when encountered, deoptimization is performed, which recovers the program state and resumes execution using bytecode.

Clearly, executing the native code after compilation is much more efficient if no deoptimization occurs.
However, when deoptimization occurs, it will take a longer time to deoptimize and roll back to the bytecode.
This difference in execution time induces the \textsc{Topti} timing side-channel
even if branches are balanced in bytecode.
When the attacker can feed inputs to the program, \textsc{Topti} could be triggered for a conditional statement whose condition relies on secrets. The attacker would then be able to infer the secret
information from the difference between execution time.

As an example, consider the \textcolor{blue}{pwdEq} method shown in Figure~\ref{fig:ToptiDemo-prog}, which is extracted and simplified from the DARPA Space/Time Analysis for Cybersecurity (STAC)
engagement program \textsl{gabfeed\_1}~\cite{STAC}. It takes strings $a$ and $b$ with length $8$ as inputs,
denoting the user-entered and correct passwords, respectively.
It checks if they are identical 
within a loop. The flag \textsf{equal} is assigned by \textcolor{darkgreen}{false} if two chars mismatch.
To balance execution time, the dummy flag  \textsf{shmequal} is introduced and assigned
by \textcolor{darkgreen}{false} if two chars match.

The \textcolor{blue}{pwdEq} method is marked as safe in STAC and would be verified as safe by the timing side-channel verification tools Blazer~\cite{AntonopoulosGHK17} and Themis~\cite{ChenFD17} which
do not consider JIT compilation.
However, it indeed is vulnerable to \textsc{Topti}.
To trigger \textsc{Topti}, we execute \textcolor{blue}{pwdEq} 50,000 times using
two strings ``PASSWORD'' and ``password''. After that,
the else-branch is replaced by the corresponding uncommon trap, so the costly deoptimization will perform later. To produce this, we use two random strings $x$ and $y$ with length 8 such that $x[0]$ is `p', $y[0]$ is not `p',  and the rest is the same.
We collect the execution time  by executing \textcolor{blue}{pwdEq} with inputs
$(x,  $``password''$)$ and  $(y,  $``password''$)$, respectively.
This mimics the process that an attacker guesses the secret data char-by-char, avoiding guessing the entire secret data simultaneously. The distribution of the execution time is shown in Figure~\ref{fig:ToptiDemo-JITenabled}.
As a cross reference, Figure~\ref{fig:ToptiDemo-JITdisabled} shows the distribution of execution time with JIT compilation disabled.
We can observe that
the difference in the execution time between two branches is much larger when JIT compilation is enabled, allowing the attacker to infer if the first char is correctly guessed.

\smallskip\noindent{\bf Branch prediction (\textsc{Tbran}).}
Branch prediction is a conservative optimization of conditional statements.
Instead of pruning rarely executed branches, branch prediction
generates native code by reordering
the basic blocks to avoid jumps 
over frequently executed branches and thus improves the spacial locality of instruction cache.
However, the reordering of basic blocks unbalances the execution time of branches even if it is balanced in bytecode.

If the attacker can feed inputs to the program,
\textsc{Tbran} could be triggered for a conditional statement whose condition relies on secrets, and thus the attacker could be able to infer secret information by measuring the execution time.
Although the difference in the execution time between branches via \textsc{Tbran} is small for
a single conditional statement, it may be amplified 
by repeated executions (e.g., enclosed in a loop).

\smallskip\noindent{\bf Method compilation (\textsc{Tmeth}).}
The most fundamental feature of JIT compilation is method compilation,
which can be triggered if a method is frequently invoked
or some backward jumps are frequently performed.
In practice, a sophisticated multi-tier compilation mechanism is adopted (in e.g., HotSpot), where
a method may be recompiled multiple times to more optimized native code to further improve the performance.
Meanwhile, during compilation, frequently invoked small methods could be inlined to speed up execution.

For a conditional statement with method invocations, if the attacker can enforce some methods in a branch to be frequently invoked in advance so that
those methods are (re)compiled or inlined, the execution time of this branch may be shortened. This difference in execution time between branches induces a timing side-channel, called \textsc{Tmeth}.

Concrete demonstrations of \textsc{Tbran} and \textsc{Tmeth} are given in the supplementary material.

\subsection{Automatically Eliminating JIT-induced Leaks}\label{sec:overview-eliminate}
In this work, we present an automated, rigorous approach to eliminate the above-mentioned JIT leaks.
We do not consider CPU-level speculative leaks~\cite{KocherHFGGHHLM019,CauligiDGTSRB20} and cache-induced leaks~\cite{AlmeidaBBDE16,DoychevK17} for which various mitigation techniques have been proposed in literature, e.g., \cite{WCGMR19,YanCS0FT18,VassenaDGCKJTS21,YuYKMTF20,HeHL21,WuGS018}.
We assume that the bytecode program is of constant-time, which can largely be achieved by existing work
(e.g.,~\cite{Agat00,MantelS15,AntonopoulosGHK17,CauligiSJBWRGBJ19,ChenFD17}).
Our goal is to protect constant-time bytecode programs from the JIT-induced leaks discussed before. In general, there are trace-based and method-based JIT compilation approaches~\cite{InoueHWN11}, and we shall focus on the latter in this work. 

A straightforward way to prevent JIT-induced leaks is to simply 
disable JIT compilation completely or JIT compilation of the chosen methods.
Indeed, \cite{brennan2020static} proposed the following three compilation strategies, named
NOJIT, DisableC2 and MExclude.
(1) The NOJIT strategy directly disables JIT compilation (e.g., both the C1 and C2 compilers in HotSpot), so no method will be JIT compiled. 
This strategy is effective and convenient to deploy, but could lead to significant performance loss. 
(2) The DisableC2 strategy only disables the 
C2 compiler instead of the entire JIT compilation,
by which the leaks induced by the 
C2 compiler  (e.g., \textsc{Topti}) can be prevented,
but not for \textsc{Tbran} or \textsc{Tmeth}.
This strategy also sacrifices the more aggressive C2 optimization and hence may suffer from performance loss.
(3) The MExclude strategy disables JIT compilation for the user-chosen methods instead of the entire program,
by which some leaks (i.e., \textsc{Tbran} and \textsc{Topti}) 
can be prevented. This strategy cannot prevent 
from \textsc{Tmeth}, but its main shortcoming is that there is no protection for  non-chosen methods, and 
it is also unclear how to choose methods to disable.
In summary, the existing compilation strategies either incur a high performance cost
or fail to prevent all the known JIT-induced leaks.

Disabling JIT compilation at the method level is indeed unnecessary.  
Essentially, we only need to ensure that secret information will not be leaked when the methods are JIT compiled or inlined.
An important observation is that secret information can only be leaked when
there is a conditional statement whose condition relies on secret data, and at least one of the following cases occurs, namely,
\begin{enumerate}
  \item (\textsc{Tmeth} leaks) a method invoked in a branch is JIT compiled or inlined;
  \item (\textsc{Tbran} leaks) the conditional statement is optimized with the branch prediction optimization; 
  \item (\textsc{Topti} leaks) the conditional statement is optimized with the optimistic optimization;  
\end{enumerate}

Based on the above observation, we propose a novel approach {\tool}, to eliminate JIT-induced leaks.
To the best of our knowledge, this is the first work to prevent all the above JIT-induced leaks
without disabling any compiler in HotSpot, which is in a sharp contrast with
the compilation strategies proposed in \cite{brennan2020static}.


In a nutshell, {\tool} automatically locates secret branch points (program points with conditional statements whose conditions rely on secret data)
by a flow-, object- and context-sensitive information flow analysis of Java bytecode~\cite{VolpanoIS96}.
The conditional statements at those secret branch points should not be optimized via branch prediction or optimistic compilations.
%
%
It then extracts all the methods invoked in those conditional statements
and identifies those methods that should not be JIT compiled or inlined.
Based on these, we put forward a fine-grained JIT compilation. 

We introduce a type system to prove the soundness of the fine-grained JIT compilation, i.e.,  
under which the resulting program is free of the aforementioned JIT-induced leaks if its bytecode version is leakage-free.
To this end, we introduce a JVM submachine and formulate
an operational semantics 
with JIT compilation. We also provide a notion of JIT-constant-time to formalize timing side-channel security of programs under JIT compilation.
We show that a constant-time program remains constant-time under
our fine-grained JIT compilation 
if the program is well-typed under our type system.
Note that our approach does not guarantee that all the identified branch points or methods are necessary,
but the precision of our approach is assured by the advanced information flow analysis and is indeed validated by experiments in Section \ref{sect:eva}.

Finally, the fine-grained JIT compilation is implemented 
by modifying HotSpot.
Our experimental results show that our approach is significantly more effective
than DisableC2 and MExclude, and is significantly more efficient
than NOJIT.

\subsection{Threat Model}
%
%
In this work, we focus on timing side-channel leaks induced by branch prediction, optimistic compilation
and method compilation. We assume that the adversary is able to influence how bytecode is JIT compiled and deoptimized by feeding inputs to programs to trigger branch prediction and optimistic compilation of chosen conditional statements, or
method compilation and deoptimization of chosen methods.
The time for JVM profiling, JIT compilation and garbage collection is not taken into account, as they are often performed in distinct threads.
We do not consider other JIT optimizations such as
constant propagation, loop unfolding and dead elimination, which are often difficult to
be controlled by the adversary at runtime. To the best of our knowledge, no existing attack leverages
these optimizations.
We do not consider CPU-level optimizations (such as speculative execution and cache) which have been studied, e.g.,~\cite{WCGMR19,YanCS0FT18,VassenaDGCKJTS21,CauligiDGTSRB20,YuYKMTF20,HeHL21}.

\section{The Language: Syntax, Semantics and Constant-Time}
In this section, we present a fragment of JVM
and formalize timing side-channel security via the notion of constant-time.

\subsection{The JVM Submachine}\label{sec:jvm-submachine-defi}
We define a fragment JVM$_{\sf JIT}$ of JVM with (conditional and unconditional) jumps, operations to manipulate the operand stack, and method calls.
Both bytecode and native code are presented in JVM$_{\sf JIT}$.
Note that this is for the sake of presentation, as our methodology is generic and could be adapted to real instruction sets of bytecode and native code.

\smallskip
\noindent
{\bf Syntax.} Let $\LVar$ (resp. $\GVar$) be the finite set of local (resp. global) variables, $\Val$ be the set of values, $\Methods$ be a finite set of methods. A program $P$ comprises a set of methods, each of which is a list of instructions taken from the instruction set in Figure~\ref{fig:instr}. All these instructions are standard except for
the instruction {\deopt\ \md}
which is used to model uncommon traps (cf. Section~\ref{sect:overview}) .

\begin{figure}[t]
  \centering
 \[
	\scalebox{0.85}{
\begin{tabular}{rcll}
\hline
 ${\sf inst}$ & ::= & \binop\ $op$ & binary operation on the operand stack\\
     	    & $\mid$ & \push\ $v$ & push value $v$ on top of the operand stack\\
     	    & $\mid$ & \pop & pop value from top of the operand stack\\
     	    & $\mid$ & \swap & swap the top two operand stack values\\
     	    & $\mid$ & \load\ $x$ & load value of $x$ onto the operand stack\\
     	    & $\mid$ & \store\ $x$ & pop and store top of the operand stack in $x$\\
     	    & $\mid$ & \get\ $y$ & load value of $y$ onto the operand stack\\
     	    & $\mid$ & \putt\ $y$ & pop and store top of the operand stack in $y$\\
     	    & $\mid$ & \ifeq\ $j$ & conditional jump\\
     	    & $\mid$ & \ifneq\ $j$ & conditional jump\\
     	    & $\mid$ & \goto\ $j$ & unconditional jump\\
     	    & $\mid$ & \invoke\ $m$ & invoke the method $m\in \Methods$\\
     	    & $\mid$ & \return & return the top value of the operand stack\\
            & $\mid$ & \deopt\ \md & deoptimize with meta data \md \\ \hline
%
\end{tabular}
}
\] \vspace*{-3mm}
  \caption{Instruction set of JVM$_{\sf JIT}$, where $x\in \LVar$ is a local variable
  and $y\in\GVar$ is a global variable}\label{fig:instr}
\end{figure}

For each method $m$, $m[i]$ denotes the instruction in $m$ at the program point $i$ and $\argv(m)$ denotes the formal arguments of $m$.
When a method is invoked, the execution starts with
the first instruction $m[0]$.
We also denote by $m[i,j]$ for $j\geq i$ the sequence of instructions $m[i]m[i+1]\cdots m[j]$.

\smallskip
\noindent
{\bf Compilation directive.}
To model method compilation with procedure inline, branch prediction and optimistic compilation
optimizations, we use (compilation) directives which specify how the method should be (re)compiled and optimized at runtime.
%
%
%
We denote by $\Dir_m$ the set of directives of the method $m$, and by $\dir(m)$ the resulting version after compilation and optimization according to the directive $\dir$.
In particular, 
we 
use $\dir_{\emptyset}\in \Dir_m$ to denote no (re)compilation.
The formal definition of directives is given in Section~\ref{sec:methodcompile}.

In general, a  method in bytecode is compiled into native code which
may be iteratively recompiled later. Thus, we assign to each method $m$ a version number $\mathcal{V}_m$,
where the bytecode has the version number $0$, and the highest version number is $\mathcal{V}_{\max}>0$.
A directive $\dir\in\Dir_m$ is invalid if $m'=\dir(m)$ and
$\mathcal{V}_{m'}> \mathcal{V}_{m}$, otherwise $\dir$ is an invalid directive.
Intuitively, the version number $\mathcal{V}_m$ indicates the optimized level of the method $m$.
JIT recompilation only uses increasingly aggressive optimization techniques, and
rolls back to the bytecode version otherwise.

\begin{figure*}[t]
  \centering
  \footnotesize
	\begin{tabular}{cccccc}
\hline \specialrule{0em}{2pt}{2pt}
         \multicolumn{2}{c}{$\dfrac{m[\pc]=\push\ v}{\langle \pc, m, \rho, \os\rangle \leadsto \langle \pc+1, m, \rho, v\cdot\os\rangle}$} &
         \multicolumn{2}{c}{$\dfrac{m[\pc]=\pop}{\langle \pc, m, \rho, v\cdot\os\rangle \leadsto \langle \pc+1, \rho, \os\rangle}$} &
		\multicolumn{2}{c}{$\dfrac{m[\pc]=\binop\ op\quad\quad v=v_1\ op\ v_2}{\langle \pc,m, \rho, v_1\cdot v_2\cdot\os\rangle \leadsto \langle \pc+1, m, \rho, v\cdot\os\rangle}$}\\ \specialrule{0em}{3pt}{3pt}
		\multicolumn{2}{c}{$\dfrac{m[\pc]=\ifeq\ j \quad\quad v= 0}{\langle \pc,m, \rho, v\cdot\os\rangle \leadsto \langle j, m, \rho, \os\rangle}$} &
        \multicolumn{2}{c}{ $\dfrac{m[\pc]=\ifeq\ j\quad\quad v\neq 0}{\langle \pc,m, \rho, v\cdot\os\rangle \leadsto \langle \pc+1,m, \rho, \os\rangle}$} &
		\multicolumn{2}{c}{$\dfrac{m[\pc]=\swap}{\langle \pc, m, \rho, v_1\cdot v_2\cdot\os\rangle \leadsto \langle \pc+1, \rho, v_2\cdot v_1\cdot\os\rangle}$} \\\specialrule{0em}{3pt}{3pt}
		\multicolumn{2}{c}{$\dfrac{m[\pc]=\ifneq\ j \quad\quad v\neq 0}{\langle \pc,m, \rho, v\cdot\os\rangle \leadsto \langle j, m, \rho, \os\rangle}$} &
          \multicolumn{2}{c}{$\dfrac{m[\pc]=\ifneq\ j\quad\quad v= 0}{\langle \pc,m, \rho, v\cdot\os\rangle \leadsto \langle \pc+1,m, \rho, \os\rangle}$} &
         \multicolumn{2}{c}{$\dfrac{m[\pc]=\store\ x\quad\quad x\in \operatorname{dom}(\rho)}{\langle \pc,m, \rho, v\cdot\os\rangle \leadsto \langle \pc+1, m,\rho[x\mapsto v], \os\rangle}$}\\\specialrule{0em}{3pt}{3pt}
       \multicolumn{2}{c}{$\dfrac{m[\pc]=\load\ x}{\langle \pc,m, \rho,\os\rangle \leadsto \langle \pc+1,m, \rho, \rho(x)\cdot\os\rangle}$} &
         \multicolumn{2}{c}{$\dfrac{m[\pc]=\goto\ j}{\langle \pc,m, \rho, \os\rangle \leadsto \langle j,m, \rho, \os\rangle}$} &
        \multicolumn{2}{c}{$\dfrac{s\leadsto s'}{(\cheap,\heap,s,\cs) \rightarrow (\cheap,\heap,s',\cs)}$ } \\\specialrule{0em}{3pt}{3pt}
       \multicolumn{3}{c}{$\dfrac{m[\pc]=\putt\ y\quad\quad y\in \operatorname{dom}(\rho) \quad\quad s=\langle \pc+1, m,\rho, \os\rangle}{(\cheap,\heap,\langle \pc,m, \rho, v\cdot\os\rangle,\cs) \rightarrow (\cheap,\heap[y\mapsto v],s,\cs)}$}&
       \multicolumn{3}{c}{$\dfrac{m[\pc]=\get\ y\quad\quad s=\langle \pc+1,m, \rho, \heap(y)\cdot\os\rangle}{(\cheap,\heap,\langle \pc,m, \rho,\os\rangle,\cs) \rightarrow (\cheap,\heap,s,\cs)}$}
        \\\specialrule{0em}{3pt}{3pt}
        \multicolumn{3}{c}{$\dfrac{m[\pc]=\return \quad\quad s=\langle \pc',m', \rho', v\cdot\os'\rangle}{(\cheap,\heap,\langle \pc,m, \rho, v\cdot\os\rangle,\langle\pc',m',\rho',\os' \rangle\cdot\cs) \rightarrow (\cheap,\heap,s,\cs)   }$ }&
         \multicolumn{3}{c}{$\dfrac{m[\pc]=\deopt\ \md \quad \mathcal{V}_m>0\quad \mathcal{O}((\cheap,\heap,\langle \pc, m, \rho, \os\rangle,\cs),\md)=(\heap',s,\cs') }{(\cheap,\heap,\langle \pc, m, \rho, \os\rangle,\cs) \rightarrow (\cheap[m\mapsto\base (m)],\heap',s,\cs'\cdot\cs)}$}
		\\\specialrule{0em}{3pt}{3pt}
	 \multicolumn{2}{c}{$\dfrac{m[\pc]=\return }{(\cheap,\heap,\langle \pc,m, \rho, v\cdot\os\rangle,\epsilon) \rightarrow (\heap,v)}$ }& 	\multicolumn{4}{c}{$\dfrac{m[\pc]=\invoke \ m' \quad  \argv(m')=x_0,\cdots,x_k \quad \dir= \dir_{\emptyset}\quad s=\langle 0,\cheap(m'), [x_0\mapsto v_0,\cdots, x_k \mapsto v_k], \epsilon\rangle}{(\cheap,\heap,\langle \pc,m, \rho, v_k\cdot\cdots\cdot v_0\cdot\os\rangle,\cs) \rightarrow_{\dir} (\cheap,\heap,s, \langle\pc+1,m,\rho,\os \rangle\cdot\cs) }$ }
		\\ \specialrule{0em}{3pt}{3pt}
	 	\multicolumn{6}{c}{$\dfrac{m[\pc]=\invoke \ m' \quad\quad  \argv(m')=x_0,\cdots,x_k \quad\quad \dir\in\Dir_m\quad\quad \dir\neq\dir_\emptyset\quad\quad m''=\dir(m') \quad\quad  \mathcal{V}_{m''}> \mathcal{V}_{m'}}{(\cheap,\heap,\langle \pc,m, \rho, v_k\cdot\cdots\cdot v_0\cdot\os\rangle,\cs) \rightarrow_{\dir} (\cheap[m'\mapsto m''],\heap,\langle 0,m'', [x_0\mapsto v_0,\cdots, x_k \mapsto v_k], \epsilon\rangle, \langle\pc+1,m,\rho,\os \rangle\cdot\cs) }$ }
		\\ \specialrule{0em}{2pt}{2pt}
\hline
	\end{tabular}
 \vspace*{-2mm} \caption{Operational semantics of JVM$_{\sf JIT}$, where $\operatorname{dom}(\rho)$ denotes the domain of the partial function $\rho$}
 \label{fig:semantics}
 \vspace*{-1mm}
\end{figure*}

\smallskip
\noindent
{\bf State and configuration.}  A state is a tuple $\langle \pc, m, \rho, \os\rangle$ where
\begin{itemize}
	\item $\pc\in \mathbb{N}$ is the program counter that points to the next instruction in $m$; 
    \item $m\in\Methods$ is the current executing method;
    \item $\rho: \LVar\rightarrow \Val$ is a partial function from local variables to values;
	\item $\os\in \Val^*$ is the operand stack.
\end{itemize}
We denote by $\statejvm$ the set of states.
For each function $f:X\rightarrow V$, variable $x\in X$ and value $v\in V$,
let $f[x\mapsto v]$ be the function where for every $x'\in X$,
$f[x\mapsto v](x')=f(x')$ if $x'\neq x$, and $f[x\mapsto v](x')=v$ otherwise.
For two operand stacks $\os_1,\os_2\in \Val^*$,
let $\os_1\cdot\os_2$ denote their concatenation. 
The empty operand stack is  denoted by $\epsilon$.


%

A configuration is of the form $(\cheap, \heap,s,\cs)$ or $(\heap,v)$, where
$\cheap$ is a code heap storing the latest version of each method;
$\heap: \GVar\rightarrow \Val$ is a (data) heap, i.e., a partial function from global variables to values;
$s\in \statejvm$ is the current state; $\cs\in \statejvm^*$
is the call stack, and $v\in \Val$ is a value.
Configurations of the form $(\heap,v)$ are final configurations, reached after the return
of the entry point.
A configuration $(\cheap,\heap,\langle \pc,m, \rho, \os\rangle,\cs)$ is an initial one if
$\pc=0$, $m$ is the entry point of the program, and $\os=\cs=\epsilon$.
%
Let $\Conf$ denote the set of configurations,
$\cs_1\cdot\cs_2$ be the concatenation of two call stacks
$\cs_1$ and $\cs_2$, and $\epsilon$ be the empty call stack. 

\smallskip
\noindent
{\bf Operational semantics with JIT Compilation.}
The small-step operational semantics of JVM$_{\sf JIT}$ is given
in Figure~\ref{fig:semantics} as a relation $\rightarrow\subseteq \Conf\times \Conf$,
where $\leadsto\subseteq \statejvm\times \statejvm$ is an auxiliary relation.
Directives $\dir$ apply to method invocations only, thus are associated to the relation $\rightarrow$ only for method invocations.
%
%
The semantics of each instruction is mostly standard except for the method invocation and deoptimization.
We only explain some selected ones. Full explanation refers to the supplementary material.

Instruction $\return$ ends the execution of the current method,
returns the top value $v$ of the current operand stack,
either by pushing it on top of the operand stack of the caller
and re-executes the caller from the return site if the current method
is not the entry point, or enters a final configuration $(\heap,v)$ if the current method
is  the entry point.

Instruction $\deopt\ \md$ deoptimizes the current method and rolls back to the bytecode in the interpreted mode.
This instruction is only used in native code and inserted by JIT compilers.
Our semantics does not directly model a deoptimization implementation.
Instead, we assume there is a deoptimization oracle $\mathcal{O}$
which takes the current configuration and the meta data $\md$ as inputs,
and reconstructs the configuration (i.e.,  heap $\heap'$, state $s$ and the call stack 
$\cs'$). Furthermore, the bytecode version $\base(m)$ of the method $m$
is restored into the code heap $\cheap$. We assume that the oracle $\mathcal{O}$ results
in the same heap $\heap'$, state $s$ and call stack $\cs'\cdot\cs$
as if the method $m$ were not JIT compiled.

The semantics of method invocation $\invoke\ m'$ depends on the directive $\dir$.
If $\dir$ is $\dir_\emptyset$ 
then the  instructions of $m'$ in the code heap $\cheap$ remain the same.
%
If  $\dir$ is valid, namely,
the optimized version $\mathcal{V}_{m''}$ after applying $\dir$ has larger version number than that of the current version $\mathcal{V}_{m'}$, the new optimized version $m''=\dir(m')$
is stored in the code heap $\cheap$.
After that, it pops the top $|\argv(m')|$
values from the current operand stack, passes them to
the formal arguments $\argv(m'')$ of $m''$, pushes
the calling context on top of the call stack
and starts to execute $m''$ in the code heap.


To define a JIT-execution, 
we introduce the notion of schedules.
A valid schedule $\dir^{\star}$ for a configuration $c$ 
is a sequence of valid directives
such that the program will not get stuck when starting from $c$ and following $\dir^{\star}$ for method invocations.
The valid schedule $\dir^{\star}$ yields a
JIT-execution, denoted by $c_0 \Downarrow_{\dir^{\star}} c_n$, which is a sequence
$c_0c_1\cdots c_n$, such that $c_0$ is an initial configuration, $c_n$ is the final configuration,
and for every $0\leq i <n$, either $c_i \rightarrow c_{i+1}$ or $c_i \rightarrow_{\dir_i}  c_{i+1}$.
We require that $\dir^\star$ is equal to the sequence of directives along the JIT-execution, i.e., the concatenation of $\dir_i$'s.
%
A JIT-free execution is thus a JIT-execution $c_0 \Downarrow_{\dir^{\star}_{\emptyset}} c_n$.
Note that in this work, we assume that the execution of a program
always terminates.

In the rest of this work, we assume that
each method has one return instruction which does not appear in any branch of
conditional statements, as early return often introduces timing side-channel leaks.


\subsection{JIT Optimization of JVM$_{\sf JIT}$}
In this section, we first introduce branch prediction and optimistic compilation,
then define method compilation as well as compilation directives in detail.

\subsubsection{Branch prediction.} \label{sec:branch-prediction-opti}

Consider a method $m$ and a conditional instruction $m[i]=\ifeq\ j$. 
(We take $\ifeq$ as the example, and $\ifneq$ can be handled accordingly. )
%
%
Let $B_{\tt t}$ (resp.\ $B_{\tt f}$) be the instructions appearing in the if-(resp.\ else-)branch of $m[i]$,
and the last instruction $m[i']$ of $B_{\tt f}$ is \goto\ $j'$.
The first and last instructions of $B_{\tt f}$ 
are $m[i+1]$ and $m[j-1]$ 
respectively.

If the profiling data show that the program favors the else-branch, 
the branch prediction optimization 
transforms the method $m$ into a new method $m_1$ as shown in Figure~\ref{fig:branch-predication-opti} for $m[i]=\ifeq\ j$.
The formal definition and an illustrating example are given in the supplementary material.

If the profiling data shows that the program favors the if-branch, 
the branch prediction optimization to the conditional instruction $m[i]=\ifeq\ j$ transforms
the method $m$ into a new method $m_2$, similar to $m_1$, except that
(1) the conditional instruction $\ifeq\ j$ is replaced by $\ifneq\ |m|-j +i-1$ which is immediately
followed by the if-branch $B_t$;
(2) the else-branch $B_f$ is moved to the end of the method starting at the point
$|m|-j +i-1$ and the target point of the last instruction \goto \ $j'$ is revised to $j'-j+i+1$.

We denote by $\transform_{\tt bp}(m,i,\elseb)$ and $\transform_{\tt bp}(m,i,\ifb)$ the new methods $m_1$ and $m_2$ respectively.
%
It is easy to see that the branch prediction optimization transforms the original program to a semantically equivalent program.

\begin{figure}[t]
  \centering
  \includegraphics[width=0.4\textwidth]{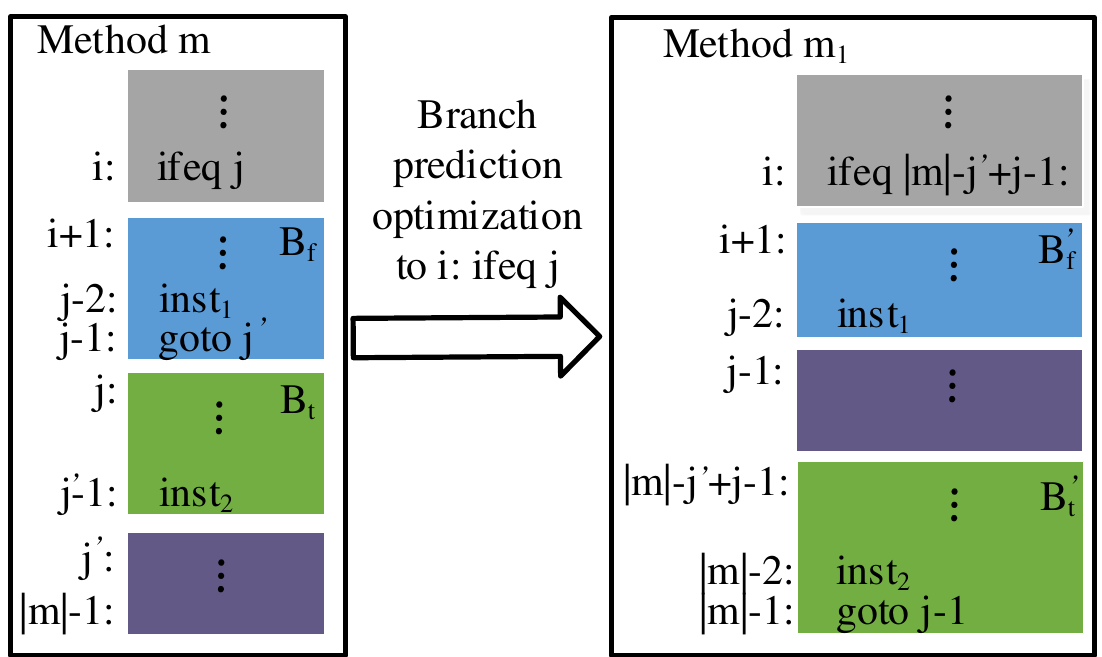}\vspace*{-2mm}
  \caption{Branch prediction optimization}\label{fig:branch-predication-opti}\vspace*{-2mm}
\end{figure}


%

\subsubsection{Optimistic compilation.}

Again, consider the conditional instruction $m[i]=\ifeq\ j$  
with the if-branch $B_{\tt t}$ and else-branch $B_{\tt f}$. ($\ifneq$ can be dealt with accordingly.)
If the profiling data show that the if-branch 
almost never gets executed,
the optimistic compilation optimization 
transforms
the method $m$ into a new method $m_1$ 
in a similar way to $\transform_{\tt bp}(m,i,\elseb)$. 
Here, the if-branch $B_{\tt t}$ 
is replaced by an uncommon trap. 
The method $m_2$ is defined similarly if the profiling data show that the else-branch 
almost never gets executed. More details refer to the supplementary material.
%

We denote by $\transform_{\tt oc}(m,i,\elseb)$
and $\transform_{\tt oc}(m,i,\ifb)$ the new methods $m_1$ and $m_2$ after transformation. 
It is easy to see that the optimistic compilation optimization is an equivalent program transformation
under the inputs that does not trigger any uncommon traps.

\subsubsection{Method Compilation}\label{sec:methodcompile}
At runtime, frequently executed, small methods 
may be inlined
to reduce the time required for method invocations.
After that, both branch prediction and optimistic compilation
optimizations could be performed. 
Thus, a compilation directive of a method
should take into account procedure inline, branch prediction and optimistic compilation
optimizations.


We define a compilation directive $\dir$ of a method $m$ as a pair
$(t,\omega)$, where $t$ is a labeled tree
specifying the method invocations to be inlined,
and $\omega$ is a sequence 
specifying the optimizations of branches.
Formally, the labeled tree $t$ is a tuple $(V, E, L)$,
where  $V$ is a finite set of nodes such that each $n\in V$ is labeled by a method $L(n)$
and the root is labeled by $m$;
 $E$ is a set of edges of the form
$(n_1,i,n_2)$ denoting that the method $L(n_2)$ is invoked at the call site $i$ of  the method $L(n_1)$.
We denote by $t(m)$ the new method obtained from $m$
by iteratively inlining method invocations in $t$.
We assume the operand stack of each inlined method is balanced,
otherwise the additional pop instructions are inserted.

The sequence $\omega$ is of the form $(\transform_1,i_1,b_1),\cdots,(\transform_k, i_k,b_k)$, where
for every $1\leq j\leq k$, $\transform_j\in \{\transform_{\tt bp},\transform_{\tt oc}\}$ denotes the optimization type
to be applied to the branch point $i_j$ in the method $t(m)$ with the branch preference $b_j$.
We assume that an index $i_j$ occurs at most once in $\omega$,
as at most one optimization can be applied to
one branch point.

Note that in our formalism, the optimistic compilation
optimization adds one uncommon trap for each conditional statement. In practice, multiple conditional statements may share one uncommon trap, which is not modeled here but can be handled by
our approach as well. 

\subsection{Consistency and Constant-Time}
We assume that each program is annotated with a set of public input variables,
while the other inputs are regarded as secret input variables.
We denote by $c_0\simeq_{\tt pub} c_0'$ if the initial configurations $c_0$ and $c_0'$ agree on the public input variables,
and denote by $c_0\simeq_{\tt ch} c_0'$ if $c_0$ and $c_0'$ have the same
code heap.

\smallskip
\noindent{\bf Consistency}.
The following theorem ensures the equivalence of the final memory store and return value
from the JIT-free execution and JIT-execution.

\begin{theorem}\label{thm:jitcorrectness}
For each initial configuration $c_0$ of the program $P$
and each valid schedule $\dir^{\star}$
for $c_0$,  we have:
\begin{center}
  $c_0 \Downarrow_{\dir^{\star}_{\emptyset}} c$ iff $c_0 \Downarrow_{\dir^{\star}} c'$.
\end{center}
\end{theorem}


If the output variables are partitioned into public and secret, we
denote by $c\simeq_{\tt pub} c'$ that the final configurations $c$ and $c'$ agree on the public output variables.


\begin{theorem}
For each pair of initial configurations $(c_0,c_0')$ of the program $P$
with $c_0\simeq_{\tt pub} c_0'$ and each pair of valid schedules $\dir^{\star}_1$ and $\dir^{\star}_2$
for $c_0$ and $c_0'$ respectively, we have: $c_0 \Downarrow c$, $c_0' \Downarrow c'$ and $c\simeq_{\tt pub} c'$
iff $c_0 \Downarrow_{\dir^{\star}_1} c$, $c_0' \Downarrow_{\dir^{\star}_2} c'$ and $c\simeq_{\tt pub} c'$.
\end{theorem}
The theorem states that observing public output variables cannot distinguish secret inputs
without JIT compilation iff observing public output variables cannot distinguish secret inputs
with JIT compilation.


\smallskip
\noindent{\bf Constant-time.}
To model execution time, we define cost functions
for bytecode and native code. Let $\cost_{\tt bc}$ and $\cost_{\tt nc}$ be the cost functions for instructions from the  bytecode and native code, respectively.
We assume that, for each pair $({\sf inst}_1,{\sf inst}_2)$ of instructions, $\cost_{\tt bc}({\sf inst}_1)=\cost_{\tt bc}({\sf inst}_2)$ implies that
$\cost_{\tt nc}({\sf inst}_1)=\cost_{\tt nc}({\sf inst}_2)$. Namely,
the cost equivalence of bytecode instructions are preserved in native code.
We denote by $\cost({\sf inst})$ the cost of the instruction ${\sf inst}$,
which is $\cost_{\tt bc}({\sf inst})$ if it is running in bytecode mode, otherwise
$\cost_{\tt nc}({\sf inst})$.
We lift the function $\cost$
to states and configurations as usual, e.g.,
$\cost(\langle \pc,m, \rho, \os\rangle)=\cost(m[\pc])$. 
The cost $\cost(c_0 \Downarrow_{\dir^{\star}} c_n)$ of a JIT-execution $c_0 \Downarrow_{\dir^{\star}} c_n$ is the sum of
all the costs of the executed instructions, i.e.,
$\sum_{i=0}^{n-1} \cost(c_i)$.

A program $P$ is constant-time (without JIT compilation)
if for each pair of initial configurations $(c_0,c_0')$ of $P$
such that $c_0\simeq_{\tt pub} c_0'$ and the code heaps of $c_0$ and $c_0'$ have the same
bytecode instructions, we have:
$$\cost(c_0 \Downarrow_{\dir^{\star}_{\emptyset}} c)=\cost(c_0' \Downarrow_{\dir^{\star}_{\emptyset}} c').$$
Intuitively, the constant-time policy
requires that two JIT-free executions have the same cost
if their public inputs are the same and code heaps have the same bytecode instructions,
thus preventing timing side-channel leaks when JIT compilation is disabled.

\smallskip
\noindent{\bf JIT-constant-time.}
To define constant-time under JIT compilation, called JIT-constant-time, we first introduce
some notations.

Consider a JIT-execution $c_0 \Downarrow_{\dir^{\star}} c_n$ and a method $m$, let $\proj_m(c_0 \Downarrow_{\dir^{\star}} c_n)$
denote the projection of the sequence of executed instructions in $c_0 \Downarrow_{\dir^{\star}} c_n$  onto
the pairs $(i,m')$ each of which consists of a program point $i$ and a version $m'$ of the method $m$.
A proper prefix $\pi$ of $\proj_m(c_0 \Downarrow_{\dir^{\star}} c_n)$
can be seen as the profiling data of the method $m$ after executing these instructions,
which determines a unique compilation directive of the method $m$ after executed $\pi$.
We leave runtime profiling abstract in order to model a large variety of
JIT compilations and use $\widehat{\pi}$ to denote the profiling data of $m$ after executed instructions $\pi$
of $m$ or its compiled versions.

Let us fix a profiler $\prof$, which provides
one compilation directive $\prof_m(\widehat{\pi})$ of a method $m$ using the profiling data $\widehat{\pi}$.
The schedule ${\dir^{\star}}$ is called a $\prof$-schedule if,
for each method $m$ and   proper prefix $\pi$ of $\proj_m(c_0\Downarrow_{\dir^{\star}}c_n)$,
the next compilation directive of $m\in \dir^{\star}$ after $\pi$
is $\prof_m(\widehat{\pi})$.

\begin{lemma}
For each pair of initial configurations $(c_0,c_0')$ of $P$
with $c_0\simeq_{\tt ch} c_0'$,
and each pair of valid $\prof$-schedules $\dir^{\star}_1$ and $\dir^{\star}_2$
for $c_0$ and $c_0'$ respectively, we have:
\begin{quote}
for every method $m$, every pair $(\pi_1,\pi_2)$ of proper prefixes  of $\proj_m(c_0 \Downarrow_{\dir^{\star}_1} c)$
and $\proj_m(c_0' \Downarrow_{\dir^{\star}_2} c')$ respectively,
if $\pi_1=\pi_2$ then $\prof_m(\widehat{\pi_1})=\prof_m(\widehat{\pi_2})$.
\end{quote}
\end{lemma}
Intuitively, the lemma ensures
that the compilation directives of each method in JIT-executions are the same under the same profiling data.

A program $P$ is JIT-constant-time
if, for every pair of initial configurations $(c_0,c_0')$ of $P$
with $c_0\simeq_{\tt pub} c_0'$ and $c_0\simeq_{\tt ch} c_0'$,
every pair of valid $\prof$-schedules $\dir^{\star}_1$ and $\dir^{\star}_2$ for $c_0$ and $c_0'$ respectively satisfies
$$\cost(c_0 \Downarrow_{\dir^{\star}_1} c)=\cost(c_0' \Downarrow_{\dir^{\star}_2} c').$$
Intuitively, the JIT-constant-time policy requires that two JIT-executions have the same cost
if their public inputs and initial code heap are the same and the valid schedules have the same profiler $\prof$ for JIT compilation, so it prevents
timing side-channel leaks even if the JIT compilation is enabled.
We allow the code heaps in $c_0$ and $c_0'$ to be mixed with bytecode and native code, because
the adversary can run the program multiple times with chosen inputs before launching attacks.

We remark that our definition of $\prof$-schedules considers a powerful adversary
who controls executing instructions and thus the compilation directives of methods, which is common in the study of detection and mitigation. In practice, the feasibility of compilation directives depends on various parameters in VM, e.g.,
whether a method invocation should be inlined depends on its code size, invocation frequency,
method modifier, etc.

As argued above, a constant-time program $P$
may not be of JIT-constant-time due to JIT compilation.
This work aims to prescribe a fine-grained JIT compilation under which a constant-time program $P$ is still JIT-constant-time
so there is no need to disable JIT compilation completely.

\begin{figure*}[t]
  \centering\setlength{\tabcolsep}{1.5pt}
  \footnotesize
	\begin{tabular}{cccccc}
\hline \specialrule{0em}{5pt}{5pt}
 \multicolumn{2}{c}{$\dfrac{m[i]= \push\ v \quad \myhl{\st'=\pt\cdot \st}}{m,i \vdash (\pt,\heapt,\lt,\st)\Rightarrow (\pt,\heapt,\lt,\st') }${\sc T-Push}}  &
 \multicolumn{2}{c}{$\dfrac{m[i]=\binop\ op  \quad \myhl{\st'=(\tau_1\sqcup\tau_2\sqcup\pt)\cdot\st}}{m,i \vdash  (\pt,\heapt,\lt,\tau_1\cdot \tau_2\cdot \st)\Rightarrow (\pt,\heapt,\lt,\st') }${\sc T-Bop}} &
 \multicolumn{2}{c}{$\dfrac{m[i]= \store\ x \quad \myhl{\lt'=\lt[x\mapsto\tau \sqcup \pt]}}{m,i \vdash (\pt,\heapt,\lt,\tau\cdot \st)\Rightarrow (\pt,\heapt,\lt',\st) }${\sc T-Str}}
 \\ \specialrule{0em}{5pt}{5pt}
 \multicolumn{2}{c}{$\dfrac{m[i]= \pop  \quad \myhl{\st=\tau\cdot \st'}}{m,i \vdash (\pt,\heapt,\lt,\st)\Rightarrow (\pt,\heapt,\lt,\st') }${\sc T-Pop}} &
 \multicolumn{2}{c}{$\dfrac{m[i]=\swap  \quad \myhl{\tau_1'=\tau_1\sqcup\pt} \quad \myhl{\tau_2'=\tau_2\sqcup\pt}}{m,i \vdash  (\pt,\heapt,\lt,\tau_1\cdot \tau_2\cdot \st )\Rightarrow (\pt,\heapt,\lt,\tau_2'\cdot \tau_1'\cdot \st) }${\sc T-Swap}} &
 \multicolumn{2}{c}{$\dfrac{m[i]= \load\ x \quad  \myhl{\st'=(\lt(x)\sqcup \pt)\cdot \st}}{m,i \vdash (\pt,\heapt,\lt,\st)\Rightarrow (\pt,\heapt,\lt,\st') }${\sc T-Load}}
 \\ \specialrule{0em}{5pt}{5pt}
  \multicolumn{2}{c}{$\dfrac{m[i]= \putt\ y \quad \myhl{\heapt'=\heapt[y\mapsto\tau \sqcup \pt]}}{m,i \vdash (\pt,\heapt,\lt,\tau\cdot \st)\Rightarrow (\pt,\heapt',\lt,\st) }${\sc T-Put}} &
   \multicolumn{2}{c}{$\dfrac{m[i]= \ifeq\ j \quad \myhl{\pt'=\tau\sqcup \pt} \quad \myhl{\pt'=\high\rightarrow i\in \prot_2(m)}}{m,i \vdash (\pt,\heapt,\lt,\tau\cdot \st)\Rightarrow (\pt',\heapt,\lt,\st) }${\sc T-If}} &
  \multicolumn{2}{c}{$\dfrac{m[i]=\goto \ j }{m,i \vdash  (\pt,\heapt,\lt,\st )\Rightarrow (\pt,\heapt,\lt, \st) }${\sc T-Goto}}
  \\   \specialrule{0em}{5pt}{5pt}
 \multicolumn{2}{c}{$\dfrac{m[i]= \get\ y \quad   \myhl{\st'= (\heapt(x)\sqcup \pt)\cdot\st}}{m,i \vdash (\pt,\heapt,\lt,\st)\Rightarrow (\pt,\heapt,\lt,\st') }${\sc T-Get}}  &
 \multicolumn{2}{c}{$\dfrac{m[i]= \ifneq\ j \quad \myhl{\pt'=\tau\sqcup \pt} \quad \myhl{\pt'=\high\rightarrow i\in \prot_2(m)}}{m,i \vdash (\pt,\heapt,\lt,\tau\cdot \st)\Rightarrow (\pt',\heapt,\lt,\st) }${\sc T-Ifn}} &
  \multicolumn{2}{c}{$\dfrac{m[i]= \return \quad \myhl{(\heapt,\tau)\models \sig_P(m)}}{m,i \vdash (\pt,\heapt,\lt,\tau\cdot\st)\Rightarrow (\heapt,\tau)  }${\sc T-Ret}}
  \\   \specialrule{0em}{5pt}{5pt}
  \multicolumn{6}{c}{$\dfrac{m[i]= \invoke \ m' \quad \argv(m')=x_0,\cdots,x_k  \quad (\pt_1,\heapt_1,\lt_1)\hookrightarrow_{m'}(\heapt_2,\tau)\quad \myhl{\pt\sqsubseteq\pt_1} \quad \myhl{\heapt\sqsubseteq \heapt_1} \quad
  \myhl{\tau_0 \sqsubseteq\lt_1(x_0)\cdots \tau_k \sqsubseteq\lt_1(x_k)} \quad \myhl{\tau'=\tau\sqcup\pt}}{m,i \vdash (\pt,\heapt,\lt,\tau_k\cdot\cdots\cdot \tau_0\cdot\st)\Rightarrow (\pt,\heapt_2,\lt,\tau'\cdot\st) }${\sc T-Call}}	
  \\ \specialrule{0em}{5pt}{5pt}\hline
	\end{tabular}\vspace*{-1mm}
\caption{Typing rules}
\label{fig:typingrules}
\end{figure*}

\section{Protect Mechanism and Type System}\label{sec:type-system}
%
%
%
%
%
%
%
%
In this section, we first propose a two-level protection mechanism to eliminate
JIT-induced leaks and then present an information-flow type system for proving
JIT-constant-time under our protected JIT compilation.

\subsection{Protection Mechanism}

The first level of our protection mechanism is to disable JIT compilation and inlining of methods
which potentially induce leaks. We denote by
$\prot_1$ the set of methods that cannot be JIT compiled or inlined, i.e.,
these methods  can only be executed in the interpreted mode.
The second level is to disable JIT optimization of branch points in methods  $\Methods\setminus \prot_1$,
whose optimization will potentially induce leaks.
We denote by $\prot_2$ the mapping from
$\Methods\setminus \prot_1$ to sets of branch  points that cannot be JIT optimized.
When the method $m$ is compiled, $\prot_2(m)$
will be updated accordingly.

From the perspective of the JVM$_{\sf JIT}$ semantics,
the compilation directive of any method
from $\prot_1$ is limited to $\dir_\emptyset$,
and the compilation directives of any method
$m'\in \Methods\setminus \prot_1$ can neither inline
a method from $\prot_1$ nor optimize the branch at a program point
in $\prot_2(m')$.
%
%


For a given program $P$, a policy for fine-grained JIT compilation is given by
a pair $(\prot_1,\prot_2)$.
%
A $\prof$-schedule ${\dir^{\star}}$ that is compliant to the policy $(\prot_1,\prot_2)$ is called a
$(\prot_1,\prot_2)$-schedule.

%
%

\subsection{Type System and Inference}
We propose an information-flow type system for proving that constant-time programs are JIT-constant-time
under a fine-grained JIT compilation with a policy $(\prot_1,\prot_2)$.

\smallskip
\noindent
{\bf Lattice for security levels.}
We consider a lattice of security levels $\seclev=\{\high,\low\}$
with $\low \sqsubseteq\low$, $\low \sqsubseteq\high$, $\high \sqsubseteq\high$
and $\high \not\sqsubseteq\low$. Initially, all the public inputs
have the low security level $\low$ and the other inputs have the high security level $\high$.
We denote by $\tau_1\sqcup\tau_2$ the least upper bound of
two security levels $\tau_1,\tau_2\in\seclev$, namely,
$\tau\sqcup \high = \high\sqcup \tau=\high$ for $\tau\in\seclev$
and $\low\sqcup \low=\low$.

\smallskip
\noindent
{\bf Typing judgments.}
Our type system supports programs whose control flow depends on secrets.
Thus, the typing rules for instructions rely on its path context $\pt$, which can indicates whether an instruction is contained in a secret branch.
We use functions $\heapt:\GVar\rightarrow \seclev$ and
$\lt:\LVar\rightarrow \seclev$ which map global
and local variables to security levels.
We also use a stack type (i.e., a stack of security levels) $\st$
for typing operand stack. The order $\sqsubseteq$ is lifted to the functions and the stack type as usual,
e.g., $\heapt_1\sqsubseteq\heapt_2$ if $\heapt_1(y)\sqsubseteq\heapt_2(y)$ for each $y\in\GVar$.

The typing judgment for non-return instructions is of the form
$m,i \vdash (\pt_1,\heapt_1,\lt_1,\st_1)\Rightarrow (\pt_2,\heapt_2,\lt_2,\st_2),$
where $m$ is the method under typing, $i$ is a program point
in $m$. 
This judgment states that, given the typing context $(\pt_1,\heapt_1,\lt_1,\st_1)$,
the instruction $m[i]$ yields a new typing context $(\pt_2,\heapt_2,\lt_2,\st_2)$.
The typing judgment of the return  is of the form
$m,i \vdash (\pt,\heapt,\lt,\st)\Rightarrow (\heapt,\tau),$
where $\heapt$ is the security levels of the global variables
and $\tau$ is the security level of the return value.

A security environment $\se_m$ of a method $m$ is a function where
for every program point $i$ of $m$, $\se_m(i)$ is a typing context
$(\heapt,\tau)$ if $m[i]$ is a return instruction, and $(\pt,\heapt,\lt,\st)$ otherwise.

\smallskip
\noindent
{\bf Method signature.}
A (security) signature of a method $m$ is of the form
$(\pt,\heapt_1,\lt_1)\hookrightarrow_{m}(\heapt_2, \tau),$
which states that, given the typing context $(\pt,\heapt_1,\lt_1)$,
each global variable $y\in \GVar$ has the security level $\heapt_2(y)$
and the return value of the method $m$ has the security level $\tau$.
Each invocation of $m$ should respect the signature of $m$.
The signature of the program $P$, denoted by $\sig_P$, is a mapping from the methods of the program $P$ to their signatures.
Since a method invoked in any secret branch cannot be JIT compiled or inlined, we require that, for any $m\in\Methods$,
$m\in\prot_1$ if the path context $\pt$ in $\sig_P(m)$ is the high security level $\high$.

\smallskip
\noindent
{\bf Typing rules.}
The typing rules are presented in Figure~\ref{fig:typingrules}, where
the key premises are \hl{highlighted} and
$(\heapt, \tau)\models \sig_P(m)$ means that
$\heapt\sqsubseteq \heapt'$ and $\tau\sqsubseteq\tau'$
for the signature $\sig_P(m)=(\pt,\heapt_1,\lt_1)\hookrightarrow_{m}(\heapt', \tau')$.

The type system only checks bytecode programs, thus
there is no typing rule for the deoptimization instruction \deopt\ \md.
Most typing rules are standard. For example, ({\sc T-Push}), ({\sc T-Pop}), ({\sc T-Bop}) and ({\sc T-Swap})
track the flow of the secret data via the operand stack, including
explicit and implicit flows.
Similarly,  ({\sc T-Str}), ({\sc T-Load}), ({\sc T-Put}) and ({\sc T-Get})
track the flow of the secret data via local and global variables.
Rule ({\sc T-Goto}) does not change the typing context.


Rules ({\sc T-If}) and ({\sc T-Ifn}) require that
the path context $\pt'$ of each branch has
a security level no less than the current
path context and the security level of the branching condition on top of the stack.
This allows us to track implicit flows during typing.
Furthermore, the branch point $i$ should not be optimized
by requiring $i\in \prot_2(m)$ if $\pt'$ has the high security level $\high$,
otherwise the branches may become unbalanced, resulting in JIT-induced leaks.

Rule ({\sc T-Ret}) requires $(\heapt, \tau)\models \sig_P(m)$ that
avoids the security levels of the global variables in $\heapt$
and the security level $\tau$ of the return value
are greater than these in the method signature $\sig_P(m)$.

Rule ({\sc T-Call}) ensures that the context of $\invoke \ m'$
meets the signature $\sig_P(m')=(\pt_1,\heapt_1,\lt_1)\hookrightarrow_{m'}(\heapt_2, \tau)$, e.g.,
$\pt\sqsubseteq\pt_1$ avoiding that the current path context $\pt$ has a security level greater than the excepted one
$\pt_1$, and $\tau_0 \sqsubseteq\lt_1(x_0)\cdots \tau_k \sqsubseteq\lt_1(x_k)$ avoiding
that actual arguments have the security levels greater than that of formal arguments.

\smallskip
\noindent
{\bf Typable methods.}
The security of a constant-time program under JIT compilation
is verified by type inference. To formalize this, we first introduce some notations \cite{BarthePR13}.

Consider a method $m$, for each program point $i$, let $\nextt_m(i)$ be
the set of successors of $i$ w.r.t.\ the control flow. Formally,
$\nextt_m(i)=\{j\}$ if $m[i]$ is $\goto\ j$,
$\nextt_m(i)=\{i+1,j\}$ if $m[i]$ is $\ifeq\ j$ or $\ifneq\ j$,
$\nextt_m(i)=\emptyset$ if $m[i]$ is $\return$,
and $\nextt_m(i)=\{i+1\}$ otherwise.

For each branch  point $i$, let $\junc(i)$ denote its junction point,
i.e., the immediate post-dominator of $i$. (Recall that
we assumed there is no early return in branches, thus $\junc(i)$ is well-defined.)
We denote by $\region(i)$ the set of program points $j$ that can be reached from the branch point $i$
and are post-dominated by $\junc(i)$. 
%
We denote by $\maxbp(j)$ the set of branch points $i$
such that $j=\junc(i)$ and $\region(i)\not\subset\region(i')$ for any $i'\in\maxbp(j)$.
Intuitively, $\maxbp(j)$ contains the branch points $i$ with
the junction point $j$ and  $\region(i)$ is not contained by
$\region(i')$ of any other branch point $i'$ with the same junction point
$j$, namely, nested branch  points $i'$ of the branch point $i$
are excluded.

A method $m$ is typable w.r.t. the signature $\sig_P$
and policy $(\prot_1,\prot_2)$, denoted by $(\prot_1,\prot_2,\sig_P)\rhd m$, if
there exists a security environment $\se_m$ for $m$ such that
$\se_m(0)=(\pt,\heapt,\lt,\epsilon)$ for  $\sig_P(m)=(\pt,\heapt,\lt)\hookrightarrow_{m}(\heapt', \tau)$
and one of the following conditions holds for each program point $i$:
\begin{itemize}
  \item if $i$ is not a junction point, then $m,j \vdash \se_m(j)\Rightarrow \se_m(i)$ for the program point $j$ such that $\nextt_m(j)=\{i\}$;
  \item if $i$ is a junction point, suppose $\se_m(i)=(\pt,\heapt,\lt,\st)$, then the following
  two conditions hold:
 \begin{itemize}
  \item there exists some $j\in \maxbp(i)$ with $\pt'\sqsubseteq \pt$ and $\se_m(j)=(\pt',\heapt',\lt',\st')$;
  \item $\heapt\sqsubseteq \heapt'$, $\lt\sqsubseteq \lt'$ and $\st\sqsubseteq \st'$ for $\nextt(j)=i$ and $\se_m(j)=(\pt',\heapt',\lt',\st')$.
\end{itemize}
\end{itemize}

Intuitively,  $(\prot_1,\prot_2,\sig_P)\rhd m$ requires
that (1) secret branches are forbidden to be optimized
by $\prot_2$ and (2) methods $m'$ invoked in $\region(i)$  of any secret branches $m[i]$
are forbidden to be JIT compiled and inlined.
Recall that we have assumed $m'\in\prot_1$ if the path context $\pt$ in $\sig_P(m')$ has the high security level $\high$.

A program $P$ is typable w.r.t. the signature $\sig_P$
and policy $(\prot_1,\prot_2)$, denoted by $(\prot_1,\prot_2,\sig_P)\rhd P$, if
(1) for the entry point $m$: $\sig_{m}=(\low,\heapt,\lt)\hookrightarrow_{m}(\heapt',\tau)$,
$\heapt(y)=\high$ and $\lt(x)=\high$ for any secret inputs $x,y$;
and (2) $(\prot_1,\prot_2,\sig_P)\rhd m$ for every method $m\in\Methods$.

\begin{theorem}\label{thm:soundness}
Given a program $P$, if $P$ is constant-time and $(\prot_1,\prot_2,\sig_P)\rhd P$,
then $P$ is JIT-constant-time under $(\prot_1,\prot_2)$-schedules.
\end{theorem}

The proof is provided in the supplementary material.
Note that the native code in the code heap of each initial configuration
can only be complied from bytecode following the policy $(\prot_1,\prot_2)$.


\vspace{-1mm}

\section{Implementation}\label{sec:implementation}

We have implemented the detection and elimination approach as a tool {\tool} for real-world Java bytecode (in the form of Jar packages).
{\tool} consists of two main components: type inference for computing
a signature $\sig_P$ and a policy $(\prot_1,\prot_2)$ such that $(\prot_1,\prot_2,\sig_P)\rhd P$, and a modified version of HotSpot from OpenJDK~\cite{OpenJDK} implementing the
protection mechanism.


\subsection{Type Inference}
Our type inference is built on JOANA~\cite{HammerS09}, a sound, flow-, context-, and object-sensitive information flow framework based on program dependence graphs.

Given a program $P$ annotated with public inputs, we first identify secret inputs and then leverage JOANA to compute a security environment $\se_m$
and a signature $\sig_P(m)$ for each method $m$ via solving flow equations.
With $\se_m$ and signature $\sig_P(m)$,
we can locate all the branch points in each method $m$ whose path context or branching condition has the high security level
$\high$, namely, all the secret branches.
These branch points are added in $\prot_2(m)$, as they can potentially induce \textsc{Topti}
and \textsc{Tbran} leakage when optimized.

From the branch points $\prot_2(m)$, we identify and extract all the methods
invoked within $\region(i)$ for all the branch points $i\in \prot_2(m)$.
These methods can potentially induce \textsc{Tmeth} leakage
when JIT compiled or inlined. Thus,
these methods are added in  $\prot_1$.
According to our type system, the soundness and precision of our type inference inherit from
that of JOANA, namely, the program $P$ is typable
w.r.t. the signature $\sig_P$
and policy $(\prot_1,\prot_2)$, i.e.,
$(\prot_1,\prot_2,\sig_P)\rhd P$ holds.

\subsection{Protection Mechanism in HotSpot}
To enforce the policy $(\prot_1,\prot_2)$ during JIT compilation, 
we modify HotSpot from OpenJDK to demonstrate our approach.
To prevent a method $m\in\prot_1$ from being compiled and inlined, we use the option \texttt{CompileCommand} supported
by HotSpot~\cite{HotSpot21}, namely,
\begin{center}\small
  -XX:CompileCommand=exclude, signature\_of\_the\_method \\
  -XX:CompileCommand=dontinline, signature\_of\_the\_method
\end{center}
where the option \texttt{exclude} disables  JIT compilation of the method \texttt{signature\_of\_the\_method},
and \texttt{dontinline} prevents the method \texttt{signature\_of\_the\_method} from procedure inline.

Unfortunately, HotSpot does not provide any option that can be used to specify branch points where branch prediction and/or optimistic compilation
can be disabled. Therefore, we modified HotSpot 
to support an additional command \texttt{dontprune}
that allows us to specify branch points. 
The command \texttt{dontprune} is used similar to \texttt{exclude}, but
with an additional list of branch points for the specified method.
During  JIT compilation, both branch prediction and optimistic compilation are prohibited for all these branch points,
even the method is recompiled. We plan to create a pull request of our modification to OpenJDK.

\subsection{\toollight}

In the experiments, we found that disabling JIT compilation of all the methods invoked in secret branches may degrade the performance significantly. To compensate, we propose and implement an alternative protection mechanism  \toollight.

{\toollight} only disables the inlining of the methods $m\in\prot_1$
whereas {\tool} disables both JIT compilation and inlining of the methods $m\in\prot_1$.
This weaker protection mechanism is still sound under the assumption that the methods invoked on both sides of each secret branch point are the same.
This assumption is reasonable in practice, as it is a straightforward strategy for developers
to implement a constant-time program by invoking same methods in both sides of each secret branch point.

We remark that inlining method should be disabled even if this method is invoked on both sides of a secret branch point,
as the method may be inlined only in one branch,
inducing branch unbalance and subsequent leakages.

%

\vspace{-1mm}

\section{Evaluation} \label{sect:eva}

In this section, we report the evaluation of {\tool} and {\toollight}.
We first evaluate the efficiency of the type inference,
and then 
compare our protection approach with other strategies: NOJIT, DisableC2, and MExclude (cf.\ Section~\ref{sec:overview-eliminate}).
According to \cite{brennan2020static}, we disable JIT compilation of the methods
that contain some secret branch points for MExclude, but methods invoked
in secret branches could be JIT complied or inlined.
%



\begin{table}[t]
  \centering \setlength{\tabcolsep}{6pt}
  \caption{Results of type checking}\vspace*{-3mm}
  \label{tab:efficiency}
  \scalebox{0.8}{
      \begin{tabular}{c|cc|cc}
      \toprule
     & {\bf Name}  &   {\bf $\sharp$LOC} & {\bf Time (s)}    &  {\bf Memory (Mb)}    \\ \midrule\rowcolor{gray!20}  \cellcolor{white}
   &   clear            & 38     & 1.50                       & 307       \\
    &  md5              & 65     & 1.53                       & 305       \\ \rowcolor{gray!20}  \cellcolor{white}
    &  salted           & 82     & 1.59                       & 309        \\
    &  stringutils      & 20     & 1.69                       & 146         \\ \rowcolor{gray!20}  \cellcolor{white}
     \multirow{-5}{*}{\rotatebox{90}{\bf DifFuzz}}  &  authmreloaded    & 76     & 0.95     & 246          \\      \midrule
   &   array            & 17     & 0.98                        & 244         \\ \rowcolor{gray!20}  \cellcolor{white}
   &   gpt14            & 12     & 2.18                       & 208        \\
   &   k96              & 24     & 2.03                       & 220          \\ \rowcolor{gray!20} \cellcolor{white}
   &   login            & 18     & 0.92                        & 239           \\
   &   loopbranch       & 23     & 0.92                        & 232           \\  \rowcolor{gray!20} \cellcolor{white}
   &   modpow1          & 22     & 2.00                       & 227           \\
   &   modpow2          & 14     & 1.95                       & 198         \\  \rowcolor{gray!20} \cellcolor{white}
   &   passwordEq       & 18     & 1.57                       & 129           \\
   &   sanity           & 15     & 0.90                        & 236            \\ \rowcolor{gray!20} \cellcolor{white}
   &   straightline     & 13     & 1.04                       & 241          \\
   \multirow{-11}{*}{\rotatebox{90}{\bf Blazer}}   &   unixlogin        & 36     & 1.19                       & 273           \\ \midrule \rowcolor{gray!20} \cellcolor{white}
    &  bootauth         & 112    & 3.68                       & 402                 \\
    &  jdk              & 13     & 0.92                        & 233                \\ \rowcolor{gray!20} \cellcolor{white}
    &  jetty            & 14     & 1.51                       & 306               \\
    &  orientdb         & 198    & 5.65                       & 539                \\ \rowcolor{gray!20} \cellcolor{white}
    &  picketbox        & 39     & 1.51                       & 130              \\
    \multirow{-6}{*}{\rotatebox{90}{\bf Themis}}  &  spring    & 25     & 1.80                       & 150              \\  \bottomrule
      \end{tabular}}\vspace*{-3mm}
\end{table}

We conduct experiments on the benchmarks that have been used to evaluate DifFuzz \cite{nilizadeh2019diffuzz}, Blazer \cite{AntonopoulosGHK17} and Themis \cite{ChenFD17},
including real-world programs from well-known Java applications such as Apache FtpServer, micro-benchmarks from DARPA STAC
and classic examples from the literature~\cite{GenkinPT14,Kocher96,PasareanuPM16}.
Recall that we target constant-time Java bytecode.
Thus, we only consider the safe versions of the benchmarks, i.e., programs that are 
leakage-free or only have slight leaks under their leakage models without the JIT compilation.
We also exclude the benchmarks tomcat, pac4j, and tourplanner from Themis,
as tomcat and pac4j have significant leakages~\cite{BrennanSB20}
while tourplanner is time-consuming (0.5 hour per execution and we shall run each benchmark 1,000 times per branch).
The remaining benchmarks are shown in Table~\ref{tab:efficiency}, where
{\bf $\sharp$LOC} shows the number of lines in the Java source code, counted
by cloc~\cite{Cloc}. Note that for the purpose of experiments, k96*, modpow1* and modpow2* are patched versions of k96, modpow1 and modpow2,
and unixlogin is a patched version by DifFuzz to resolve the NullPointerException error in its original version from Blazer. 

All experiments are conducted on an Intel NUC running Ubuntu 18.04 with Intel Core I5-8259U CPU @ 2.30GHz and 16GB of memory,
without disabling CPU-level and other JIT optimizations when JIT compilation is enabled.

In summary, the results show that
(1) \tool\ is very effective and is able to successfully eliminate a majority of the leakages induced
by JIT compilation, and (2) {\toollight} is able to achieve comparable effectiveness as {\tool}
and induces significantly less performance loss.

\begin{table*}[t]
  \centering \setlength{\tabcolsep}{1pt}
  \caption{Evaluation results of \tool}\vspace*{-3mm}
  \label{tab:results}
  \scalebox{0.8}{
      \begin{tabular}{c|ccc*{5}{|>{\centering\arraybackslash}p{.085\linewidth}>{\centering\arraybackslash}p{.085\linewidth}}}
      \toprule
      & \multicolumn{3}{c|}{\bf Benchmark} &  \multicolumn{2}{c|}{\bf NOJIT} &  \multicolumn{2}{c|}{\bf DisableC2} &  \multicolumn{2}{c|}{\bf MExclude} &  \multicolumn{2}{c|}{\bf \tool} &  \multicolumn{2}{c}{\bf \toollight} \\ \cline{2-14}
      & {\bf Name} & {\bf Leakage} & {\bf Time (µs)}      &  {\bf Leakage} & {\bf Overhead} &  {\bf Leakage} & {\bf Overhead} &  {\bf Leakage} & {\bf Overhead} &  {\bf Leakage} & {\bf Overhead} &  {\bf Leakage} & {\bf Overhead}    \\ \midrule \rowcolor{gray!20}\cellcolor{white}
&         clear   &        1.00   &     4.846   &    0.02   &   49.40   &   0.02   &   3.47   &       0.02   &      12.95   &  \bf 0.01   &     25.22   &       1.00   &   \bf 1.00  \\
&           md5   &        1.00   &     6.526   &    0.19   &   47.81   &   0.09   &   4.13   &   \bf 0.01   &      10.00   &  \bf 0.01   &     19.51   &   \bf 0.01   &   \bf 1.82  \\\rowcolor{gray!20}\cellcolor{white}
&        salted   &        1.00   &     6.711   &\bf 0.02   &   47.80   &   0.17   &   3.93   &       0.20   &       9.69   &      0.03   &     18.99   &       0.17   &   \bf 1.77  \\
&   stringutils   &        0.97   &     0.559   &\bf 0.10   &   11.90   &   0.59   &   1.57   &       1.00   &       2.64   &      0.77   &      8.92   &       1.00   &   \bf 1.35  \\\rowcolor{gray!20}\cellcolor{white}
&   authmreloaded &        1.00   &     8.696   &\bf 0.01   &   34.89   &   0.05   &   4.46   &       0.03   &       1.28   &      0.03   &  \bf 1.00   &       0.03   &   \bf 1.00  \\ \cmidrule{2-14}
\multirow{-6}{*}{\rotatebox{90}{\bf DifFuzz}}  & {\bf Average}   &        0.99   &     5.468   &\bf 0.07   &   38.36   &   0.18   &   3.51   &       0.25   &       7.31   &      0.17   &     14.73   &       0.44   &   \bf 1.39  \\
         \midrule \rowcolor{gray!20}\cellcolor{white}
&         array   &        1.00   &      0.229   &    1.00   &    2.00   &   0.64   &   1.21   &       1.00   &       2.61   &  \bf 0.23   &  \bf 1.00   &       0.25   &   \bf 1.00  \\
&         gpt14   &        1.00   &     2.157   &\bf 0.01   &   45.11   &\bf 0.01  &   3.06   &       0.20   &       1.80   &  \bf 0.01   &     15.95   &   \bf 0.01   &   \bf 1.47  \\\rowcolor{gray!20}\cellcolor{white}
&           k96   &        1.00   &     2.414   &\bf 0.02   &   42.69   &   1.00   &   3.04   &       0.79   &       1.83   &      1.00   &     18.50   &       1.00   &   \bf 1.46  \\
&          k96*   &        1.00   &     2.372   &\bf 0.02   &   42.93   &\bf 0.02  &   3.09   &       0.59   &       1.90   &  \bf 0.02   &     18.99   &       0.52   &   \bf 1.48  \\\rowcolor{gray!20}\cellcolor{white}
&         login   &        1.00   &      0.266   &    0.79   &    2.05   &   0.67   &   1.17   &       0.91   &       2.68   &  \bf 0.54   &  \bf 1.05   &   \bf 0.54   &   \bf 1.05  \\
&    loopbranch   &        1.00   &      0.243   &    0.86   &    5.57   &   0.80   &   3.15   &       0.33   &      15.34   &  \bf 0.01   &  \bf 0.98   &   \bf 0.01   &   \bf 0.98  \\\rowcolor{gray!20}\cellcolor{white}
&       modpow1   &        1.00   &    78.615   &\bf 0.02   &    0.36   &   1.00   &   0.21   &       1.00   &       0.65   &      1.00   &  \bf 0.16   &       1.00   &       0.95  \\
&      modpow1*   &        1.00   &    78.542   &\bf 0.01   &    0.36   &   0.02   &   0.23   &       1.00   &       0.65   &  \bf 0.01   &  \bf 0.16   &   \bf 0.01   &       0.94  \\\rowcolor{gray!20}\cellcolor{white}
&       modpow2   &        1.00   &      0.789   &\bf 0.01   &   36.92   &   1.00   &   2.78   &       1.00   &       2.27   &      1.00   &     15.61   &       1.00   &   \bf 1.57  \\
&      modpow2*   &        1.00   &      0.945   &    0.01   &   42.15   &   0.07   &   2.93   &       1.00   &       2.12   &      0.01   &     17.55   &   \bf 0.00   &   \bf 1.55  \\\rowcolor{gray!20}\cellcolor{white}
&    passwordEq   &        1.00   &      0.262   &    0.13   &    6.61   &   0.17   &   1.53   &       0.56   &       3.74   &  \bf 0.01   &      5.39   &   \bf 0.01   &   \bf 1.15  \\
&        sanity   &        1.00   &      0.234   &    0.25   &    5.83   &   0.97   &   2.82   &       0.07   &      16.02   &  \bf 0.01   &  \bf 0.99   &   \bf 0.01   &       1.00  \\\rowcolor{gray!20}\cellcolor{white}
&  straightline   &        1.00   &      0.231   &    0.80   &    2.03   &   0.07   &   1.07   &       0.90   &       2.16   &  \bf 0.00   &  \bf 1.00   &       0.01   &   \bf 1.00  \\
&     unixlogin   &        1.00   &      0.316   &\bf 1.00   &    8.51   &\bf 1.00  &   1.96   &   \bf 1.00   &       3.03   &  \bf 1.00   &     10.09   &   \bf 1.00   &   \bf 1.37  \\\cmidrule{2-14}
\multirow{-12}{*}{\rotatebox{90}{\bf Blazer}}   & {\bf Average}   &        1.00   &    11.973   &\bf 0.35   &   17.37   &   0.53   &   2.02   &       0.74   &       4.06   &  \bf 0.35   &      7.67   &       0.38   &   \bf 1.21  \\
           \midrule \rowcolor{gray!20}\cellcolor{white}
&      bootauth   &        1.00   &     2.793   & 0.02   &  106.98   &   \bf 0.01   &   4.53   &       0.03   &       1.53   &      0.84   &      1.47   &       0.04   &   \bf 1.05  \\
&           jdk   &        1.00   &      0.236   &    0.16   &    2.15   &   0.05   &   1.14   &       0.19   &       2.68   &  \bf 0.01   &  \bf 1.01   &   \bf 0.01   &   \bf 1.01  \\\rowcolor{gray!20}\cellcolor{white}
&         jetty   &        1.00   &      0.254   &    0.11   &    6.49   &   0.17   &   1.51   &       0.50   &       3.51   &  \bf 0.01   &      5.48   &   \bf 0.01   &   \bf 1.14  \\
&      orientdb   &        0.99   &     1.942   &\bf 0.01   &   78.48   &   0.01   &   3.47   &       0.33   &       1.39   &  \bf 0.01   &      1.28   &   \bf 0.01   &   \bf 0.99  \\\rowcolor{gray!20}\cellcolor{white}
&     picketbox   &        1.00   &      0.252   &    0.04   &    7.23   &   0.02   &   1.54   &       1.00   &       1.82   &      0.06   &      7.85   &   \bf 0.01   &   \bf 1.30  \\
&        spring   &        1.00   &      0.509   &\bf 0.01   &   14.16   &   0.02   &   2.10   &       0.04   &       2.63   &  \bf 0.01   &      1.71   &   \bf 0.01   &   \bf 1.06  \\\cmidrule{2-14}
\multirow{-7}{*}{\rotatebox{90}{\bf Themis}} & {\bf Average}   &        1.00   &      0.998   &    0.06   &   35.92   &   0.05   &   2.38   &       0.35   &       2.26   &      0.16   &      3.13   &   \bf 0.02   &   \bf 1.09  \\
      \bottomrule
      \end{tabular}}\vspace*{-2mm}
\end{table*}

\subsection{Results of Type Inference}
Table~\ref{tab:efficiency} shows the time and memory
used for type inference of the benchmarks.
We observe that these benchmarks can be solved efficiently.
It takes 1.73 seconds on average (up to 5.65 seconds) and 251 Mb to analyze one benchmark.
Note that the time and memory consumption does not necessarily correlate with the size of the program (e.g., on \textsl{gpt14} vs. \textsl{k96}). 



\subsection{Effectiveness and Efficiency of the Protection}\label{sec:effectiveness}
Our approach provides the security guarantees of JIT-constant-time w.r.t. the JIT-induced leaks of \textsc{Topti},
\textsc{Tbran} and \textsc{Tmeth}, but there are other JIT and CPU-level optimizations that may induce timing
side-channel leaks as well. Thus, we evaluate the effectiveness
by quantifying the amount of leakages in practice using mutual information~\cite{MH10}, a widely used metric for side channel analysis \cite{KopfB07,KopfMO12,StandaertMY09,MalacariaKPPL18}.

The mutual information of a program containing a vulnerable conditional statement with the secret condition $K$ and execution time $T$ is defined as
$\mathrm{I}(K ; T)=\mathrm{H}(K)-\mathrm{H}(K|T)$, where
$\mathrm{H}(K)$ is classical Shannon entropy measuring uncertainty about
$K$, and $\mathrm{H}(K|T)$ 
is the conditional Shannon entropy of $K$ given $T$.
$\mathrm{I}(K ; T)$ measures the uncertainty about $K$ after the attacker has learned the execution time $T$.
We manually create attacks to explore the maximum amount of leakages according to~\cite{BrennanRB20}.
To discretize the execution time $T$, we split
it into a 20 bins. Note that the closer the mutual information value is to 1, the stronger the relationship between the branch condition $K$ and execution time $T$.

The results are summarized in Table \ref{tab:results}, which records the average
of 1,000 experiments for each benchmark, where the best results among different methods are in \textbf{bold face}.
The second and third columns show the leakage and execution time
without any defense.
The other columns show the leakage with the corresponding defense
and the overhead (calculated as the ratio: execution time with the defense/execution time without defense) induced by the defense.

\smallskip\noindent
{\bf Effectiveness}.
Overall, we can observe that (1) all these safe programs become vulnerable (i.e.,  nonnegligible leakage) due to JIT compilation;
(2) disabling JIT compilation (NOJIT) can effectively reduce JIT-induced leakages for most programs except for \textsl{array}, \textsl{login}, \textsl{loopbranch}, \textsl{straightline} and \textsl{unixlogin};
(3) \tool\ and \toollight\ perform significantly better than DisableC2 and MExculde, even better than NOJIT on some benchmarks (e.g., \textsl{md5}, \textsl{array}, \textsl{login}, \textsl{loopbranch}, \textsl{sanityjdk} and \textsl{jetty});
(4) \tool\ and \toollight\ are almost comparable.

\smallskip\noindent
{\bf Efficiency}. We measure the efficiency of respective method by the times the execution time is increased.
In general,
(1) NOJIT incurs the highest performance cost;
(2) DisableC2 and MExclude lead to nearly 2--7 times runtime overhead;
(3) {\tool} incurs more overhead than DisableC2 and MExclude;
(4) {\toollight} brings the least runtime overhead (up to 1.82 times).

On some benchmarks (e.g., \textsl{authmreloaded}, \textsl{array}, \textsl{login}, \textsl{loopbranch}, \textsl{sanity} and \textsl{jdk}),
{\tool} performs better than DisableC2.
It is because DisableC2 completely disables C2 mode compilation for all the methods,
whereas {\tool} disables JIT compilation and procedure inline of methods invoked in secret branches.
Thus, {\tool} performs better than DisableC2 when many methods can be compiled in the C2 mode at runtime.
We note that MExclude allows JIT compilation and inlining of methods invoked in secret branches,
thus outperforms {\tool} in general. When many methods contain secret branches but few methods are invoked therein, MExclude performs worse than  {\tool}.

We shall discuss some interesting case studies below.

\noindent {\bf array, login, loopbranch, straightline, unixlogin:}
Experimental results show that their leakages are significant in practice,
although these benchmarks were claimed of leakage-free or only have slight leaks under their leakage models without JIT compilation~\cite{nilizadeh2019diffuzz,AntonopoulosGHK17,ChenFD17}.
Interestingly, both {\tool} and {\toollight} are able to significantly reduce the leakage of \textsl{array}, \textsl{login}, \textsl{loopbranch}, and \textsl{straightline}.
This is because the percentage of timing difference is fixed, the program speeds up with the JIT compilation (i.e., lower overhead), making
side channel-unstable and difficult to observe due to the fixed noise.
The case for \textsl{unixlogin} is slightly different.
Recall that \textsl{unixlogin} is a patched version by DifFuzz to resolve the NullPointerException error in its original version from Blazer.
However, this patch introduced a leakage which is always significantly observable.

\noindent\textbf{stringutils:} We observe that only NOJIT effectively reduces the JIT-induced leakage
of \textsl{stringutils}.
We found that \textsl{stringutils} evaluates a method in Apache FtpServer that pads a string to a specified length, where an insecure version would leak information about the original string's length.
{\tool} and {\toollight} successfully eliminated the JIT-induced leak in this method, guaranteeing
the balance of secret branches in the native code. However, due to CPU-level optimizations (e.g.,  speculative execution), the execution time of different branches varies with secret inputs.


\noindent\textbf{k96, modpow1, modpow2:}
Similar to \textsl{stringutils}, we observe that only NOJIT effectively reduces their JIT-induced leakages.
These programs implement various components of the RSA cryptosystem's modular exponentiation using the classic square-and-multiply algorithm,
thus their leakages would result in key recovery attacks~\cite{Kocher96}.
{\tool} and  {\toollight} indeed can guarantee that no leaks are induced by JIT compilation in the native code.
However, due to CPU-level optimizations, the execution time of the branches varies with secret inputs. To reduce such noise, we created patched versions \textsl{k96*}, \textsl{modpow1*} and \textsl{modpow2*} by
 moving the time-consuming operations from branches to outside of their branching point.
 After patching, most defense solutions are able to reduce the JIT induced leakages.


%


\noindent\textbf{bootauth:} \tool\ is not effective on \textsl{bootauth}, due to
an unbalanced statement in bytecode.
Since \tool\ only disables JIT compilation of the methods invoked in secret branches,
but other methods can be JIT complied including the C2 mode compilation.
Thus, \tool\ is only able to eliminate the \textsc{Topti} leak,
but amplifies the timing difference of these branches compared over the entire execution time,
whereas the others do not amplify the timing difference.

\vspace{-1mm}

\section{Related Work}
%
Timing side-channel attacks have attracted many attentions, with a significant amount of work devoted to its
detection~\cite{PasareanuPM16,MalacariaKPPL18,nilizadeh2019diffuzz,BrennanSB18},
verification~\cite{DoychevFKMR13,AntonopoulosGHK17,ChenFD17,BartheBCLP14,AlmeidaBBDE16, DoychevK17, BlazyPT19} and
mitigation~\cite{Agat00,DoychevFKMR13,MantelS15,WuGS018,WattRPCS19,CauligiSJBWRGBJ19},
which vary in targeted programs, leakage models, techniques, efficiency and precision, etc.
%

More recent work focus on other sources of timing side-channels, induced by
micro-architectural features (e.g., Spectre \cite{KocherHFGGHHLM019} and Meltdown \cite{Lipp0G0HFHMKGYH18}) or compilation (e.g., JIT-induced leaks~\cite{BrennanRB20}) 
where provably leakage-free programs (or with slight leakages) may become vulnerable when they are taken into account~\cite{CauligiDGTSRB20}. Our work is within this category.


Micro-architectural features 
allow new timing side-channel attacks such as Spectre, Meltdown and  variants thereof~\cite{0001LMBS0G19,SchaikMOFMRBG19,abs-1905-12701,BulckMWGKPSWYS18,CauligiDGTSRB20}.
This problem has been recently studied~\cite{Wu019,GuarnieriKMRS20,CauligiDGTSRB20,GuoCLCWW020,GuoCYW0LCW20,WCGMR19,VassenaDGCKJTS21,YanCS0FT18,YuYKMTF20,HeHL21}, where speculative execution semantics, notions of constant-time under the new semantics,
detection and mitigation approaches, etc, have been proposed.
Among them, Blade is the closest to our work, which aims to ensure that constant-time programs are 
leakage-free under speculative and out-of-order execution.
Our work is similar in spirit, but as the leaks induced by JIT compilation and micro-architecture features are different,
the concrete technology (e.g., security notions, detection and mitigation approaches) in this paper is new. Moreover,
as discussed in our experiments, native code compiled from bytecode
may suffer from leakages induced by micro-architectural features.
Such leakages could potentially be eliminated
by integrating existing mitigation approaches (e.g., Blade)  into JIT compilation.

Besides JIT compilation, static compilation 
can also introduce
timing leakages. To address this problem, constant-time preserving compilation has been studied~\cite{BartheGL18}
and subsequently implemented in the verified compiler CompCert~\cite{BartheBGHLPT20}.
However, they disallow secret branches, increasing the difficulty of implementing constant-time programs.
Follow-up work includes constant-resource preserving compilation~\cite{BartheBHP21}.
However, neither of them considered JIT compilation which is far more complex than the static compilation.

The work on JIT-induced timing channel is very limited. The work close to ours is~\cite{BrennanRB20,BrennanSB20,brennan2020static}.
The JIT-induced leaks proposed in \cite{BrennanRB20} demonstrated how JIT compilation
can be leveraged to mount timing side-channel attacks. A fuzzing approach was proposed to detect JIT-induced leaks~\cite{BrennanSB20}.
However, it can neither prove free of JIT-induced leaks nor mitigate JIT-induced leaks.
The three strategies (i.e., NOJIT, DisableC2 and MExclude) proposed in~\cite{brennan2020static} have been discussed and compared in Section~\ref{sec:overview-eliminate} and Section~\ref{sec:effectiveness}.

\vspace{-1mm}

\section{Conclusion and Future Work}
We have presented an operational semantics and a formal definition of constant-time programs under JIT compilation,
based on which we have proposed an automated approach to eliminate JIT-induced timing
side-channel leaks. Our approach systematically detects potential
leaks via a precise information flow analysis and eliminates potential leaks
via a fine-grained JIT compilation.
We have implemented our approach in the tool {\tool}.
The evaluation shows that  {\tool} is more effective
than existing solutions and provides a trade-off between the security and performance.
The lightweight variant  {\toollight} of {\tool}  pushes the limit of overhead further
with comparable effectiveness.


In the future, we plan to improve our approach by taking into account other JIT optimizations and CPU-level optimizations.
The efficiency could also be improved by refining the granularity of our fine-grained JIT compilation, e.g., methods invoked in both branches of a secret branch point could be inlined simultaneously
which would not break the balance of the two branches.

\appendix

\begin{figure*}
  \centering
      \begin{subfigure}[b]{0.24\textwidth}
    \centering
\begin{minted}[fontsize=\footnotesize,bgcolor=bg]{java}
boolean verifyPin(int input){
    boolean equal= false;
    if (input != pin) {
        equal = false;
    } else {
        equal = true;
    }
    return r;}
  \end{minted}
  \caption{The \textcolor{blue}{verifyPin} method}
   \label{fig:TbranDemo-prog}
  \end{subfigure}
  \begin{subfigure}[b]{0.24\textwidth}
    \centering
    \includegraphics[width=\textwidth]{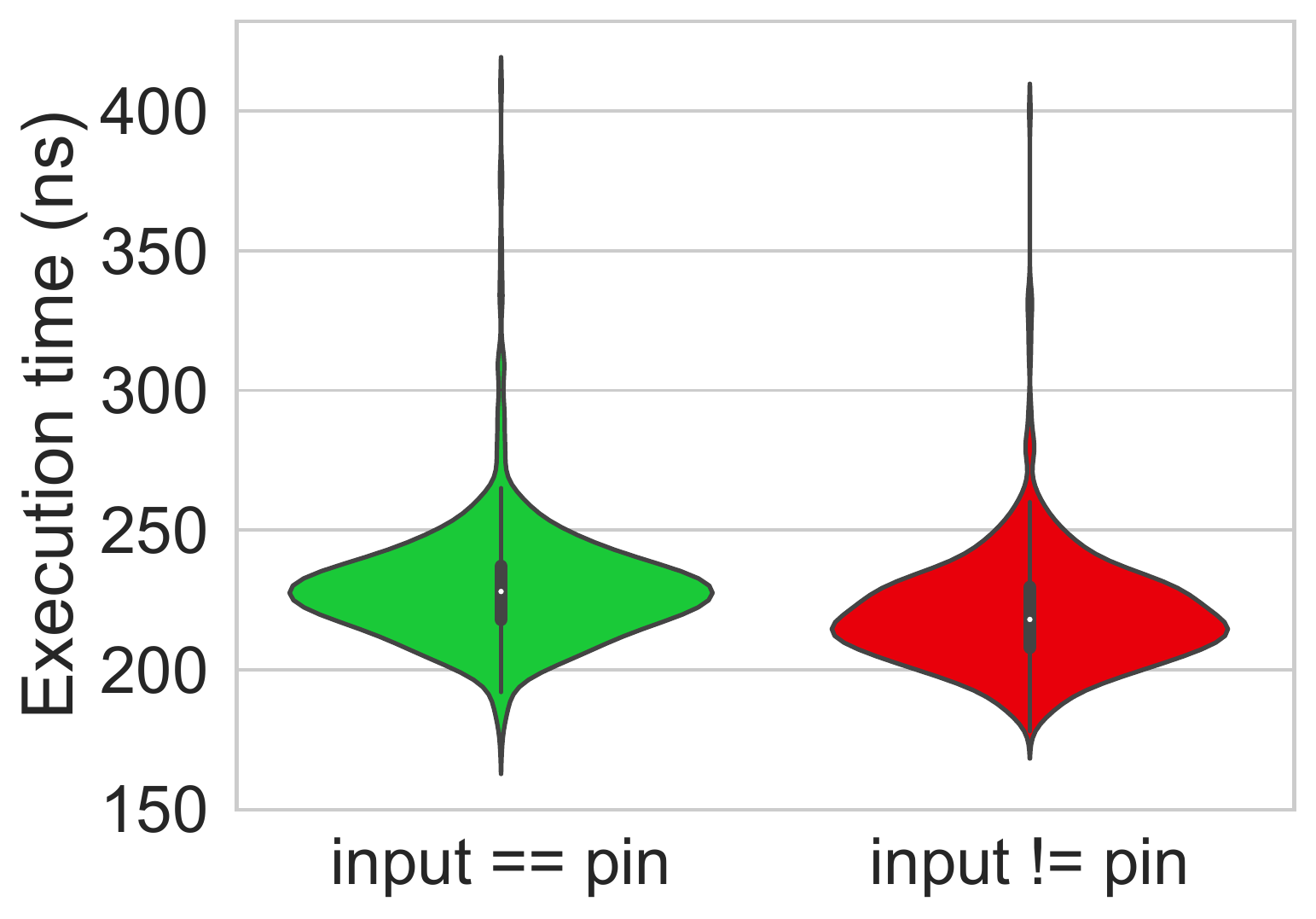}
    \caption{JIT enabled}\label{fig:TbranDemo-JITenabled}
  \end{subfigure}
    \begin{subfigure}[b]{0.24\textwidth}
    \centering
    \includegraphics[width=\textwidth]{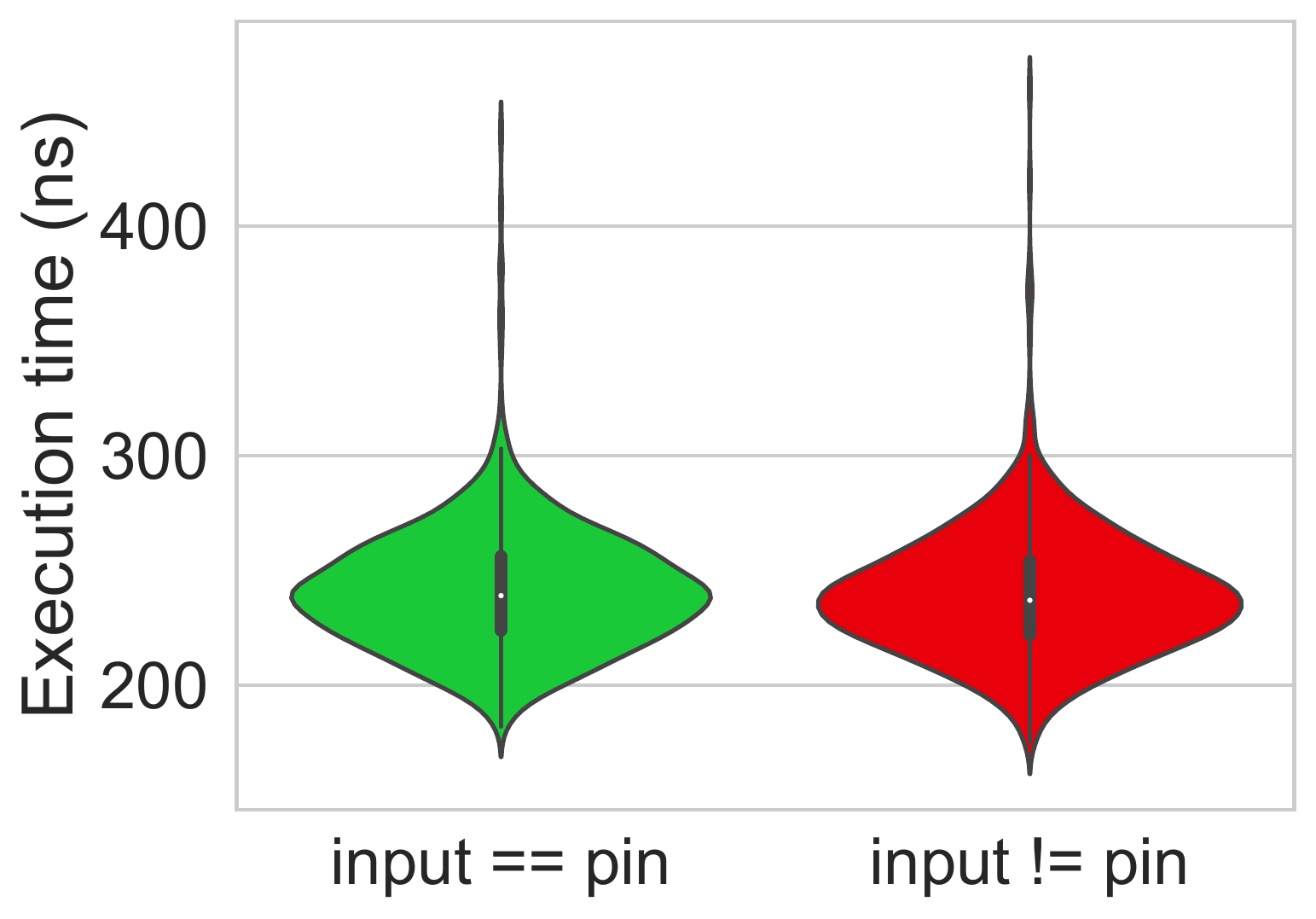}
    \caption{JIT disabled}\label{fig:TbranDemo-JITdisabled}
  \end{subfigure}
  \begin{subfigure}[b]{0.24\textwidth}
    \centering
    \includegraphics[width=\textwidth]{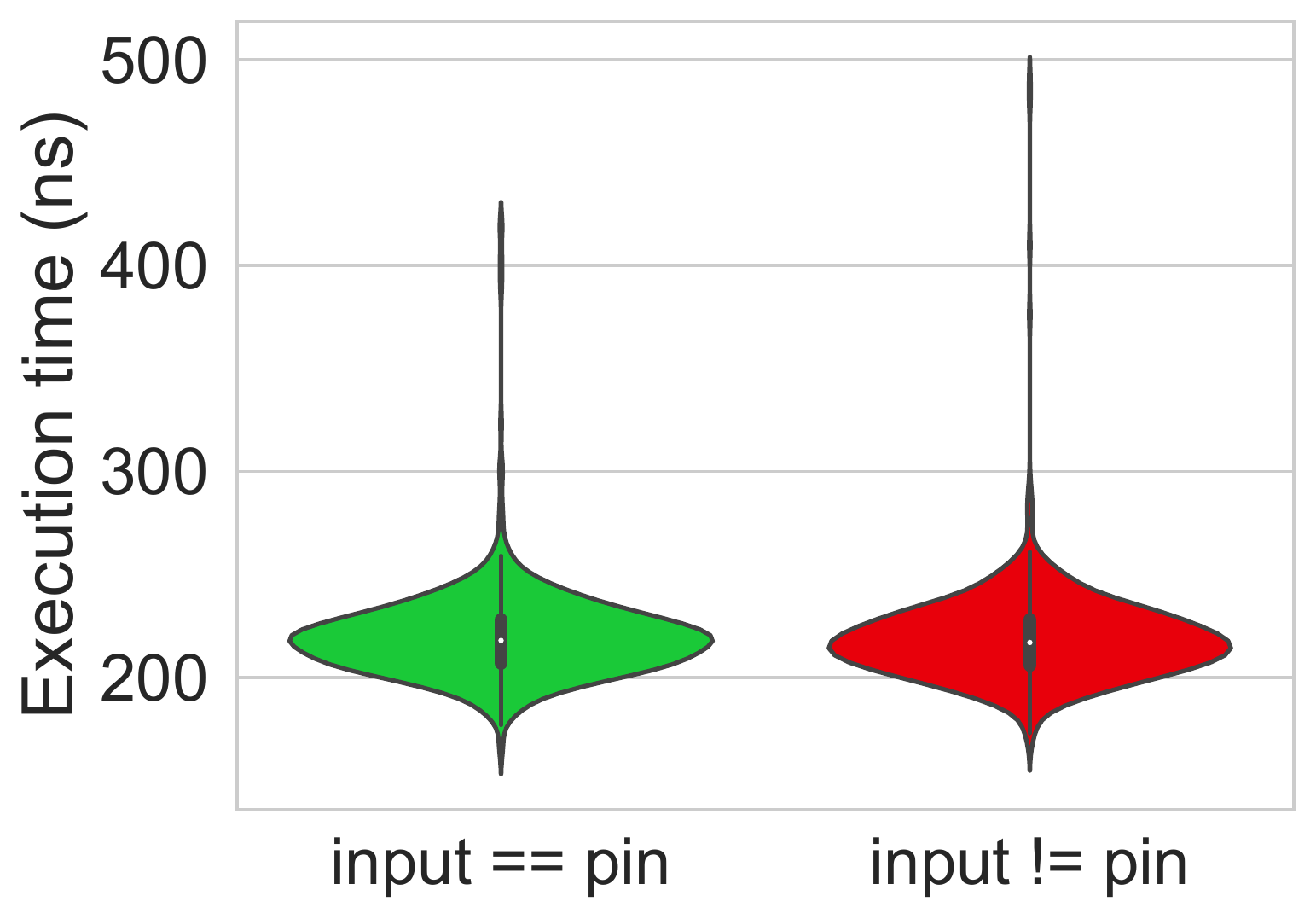}
    \caption{Mitigated}
    \label{fig:mitigationDemoverifyPin}
  \end{subfigure}
  \caption{The \textcolor{blue}{verifyPin} method and its execution time with JIT enabled and disabled under \textsc{Tbran} 
  }
  \label{fig:TbranDemo}
\end{figure*}

\section{Supplementary Material}
This is supplementary material of the submission entitled
``\tool: Eliminating  Just-In-Time Compilation Induced Timing Side-Channel Leaks".

\subsection{Demonstration of Branch Prediction (\textsc{Tbran})}\label{sec:demoofTbran}
To demonstrate the \textsc{Tbran} timing side-channel, we use a simple method \textcolor{blue}{verifyPin} shown in Figure~\ref{fig:TbranDemo-prog}.
It accepts a user-supplied parameter \textsf{input} and
 returns \textcolor{darkgreen}{false} if
 \textsf{input}!=\textsf{pin} and \textcolor{darkgreen}{true} otherwise.
  Clearly, it is not vulnerable without JIT compilation.
   However, \textcolor{blue}{verifyPin} indeed is vulnerable to \textsc{Tbran}. To trigger \textsc{Tbran}, we execute \textcolor{blue}{verifyPin} 50,000 times with \textsf{pin}=0xdeadbeaf,
   where \textsf{input} is randomly generated with probability of $\frac{1}{8}$ being 0xdeadbeaf.
The execution of the if-branch is more often than that of the else-branch, triggering the branch predication optimization. After JIT compilation,
the if-branch will take less time than the else-branch.
This is justified by executing \textcolor{blue}{verifyPin} 2,000 times
(1,000 times for \textsf{input}==\textsf{pin} and 1,000 times for \textsf{input}!=\textsf{pin}),
and computing the distribution of execution time, shown in Figure~\ref{fig:TbranDemo-JITenabled}.
  As a cross reference, Figure~\ref{fig:TbranDemo-JITdisabled} shows the distribution of execution time with JIT compilation disabled.
We can observe that the executions time of two branches is almost the same when the JIT compilation is disabled,
while there exists a gap between the execution time of two branches when the JIT compilation is enabled.
This difference allows the adversary to determine if \textsf{input} is the correct \textsf{pin} without knowing the return value of the method in a ``blind'' scenario.

By disabling the branch prediction of
the conditional statement,
the \textsc{Tbran}-induced leakage is prevented, justified by the execution time shown in
Figure~\ref{fig:mitigationDemoverifyPin}. Note that  the  \textcolor{blue}{verifyPin} method
could be JIT compiled when our approach is applied.

\subsection{Demonstration of Method Compilation (\textsc{Tmeth})}\label{sec:demoofTmeth}
To demonstrate the \textsc{Tmeth} timing side-channel, consider the \textcolor{blue}{checkSecret}  method shown in Figure~\ref{fig:TmethDemo-prog}, which is extracted and simplified
from the STAC canonical program \textsl{category1}~\cite{STAC}. It is not vulnerable in interpreted mode
when the execution time of \textsf{consume1} and \textsf{consume2} is identical.
However, if one of the methods is compiled or inlined at runtime while the another one is not,
the branches will have unbalanced execution time, thereby introducing a new timing side-channel. To justify this, we invoke  \textcolor{blue}{checkSecret}  only once with $n=500$
and \textsf{guess} being a very small number. This enforces \textsf{consume1} in the if-branch to be invoked 250,000 times, large enough to trigger the method compilation of \textsf{consume1}.
After that, as shown in Figure~\ref{fig:TmethDemo-JITenabled},  for each input \textsf{guess}, if the execution time is small, we can deduce
that \textsf{guess}$\leq$\textsf{secret}, while if the execution time is large,
we can deduce that \textsf{guess}$>$\textsf{secret}.
As a cross reference, Figure~\ref{fig:TmethDemo-JITdisabled} shows the distribution of execution time with JIT compilation disabled,
indicating that the execution time of the branches is similar.
This side-channel is very stable, as the adversary can repeatedly first invoke \textcolor{blue}{checkSecret}  in advance by passing a very small number which triggers the method compilation of \texttt{consume1},
then the subsequent invocation of \textcolor{blue}{checkSecret} uses some data to check if it is correct, leading to a leak.
To speed up the guess-and-check process, the adversary could use a binary search.

By disabling the JIT compilation of the \textsf{consume1} and \textsf{consume2} methods,
the leak induced by \textsc{Tmeth} is prevented, justified by the execution time shown in
Figure~\ref{fig:mitigationDemocheckSecret}.
Note that  the  \textcolor{blue}{checkSecret} method
could be JIT compiled when our approach is applied.

\begin{figure*}
  \centering
    \begin{subfigure}[b]{0.24\textwidth}
    \centering
 \begin{minted}[fontsize=\scriptsize,bgcolor=bg]{java}
void checkSecret(int guess) {
  if (guess <= secret) {
    for (int i = 0; i < n; i++)
      for (int t = 0; t < n; t++)
        consume1();  }
  else {
    for (int i = 0; i < n; i++)
      for (int t = 0; t < n; t++)
        consume2();  }
}
  \end{minted}
    \caption{The \textcolor{blue}{checkSecret} method}
     \label{fig:TmethDemo-prog}
   \end{subfigure}
  \begin{subfigure}[b]{0.24\textwidth}
    \centering
    \includegraphics[width=\textwidth]{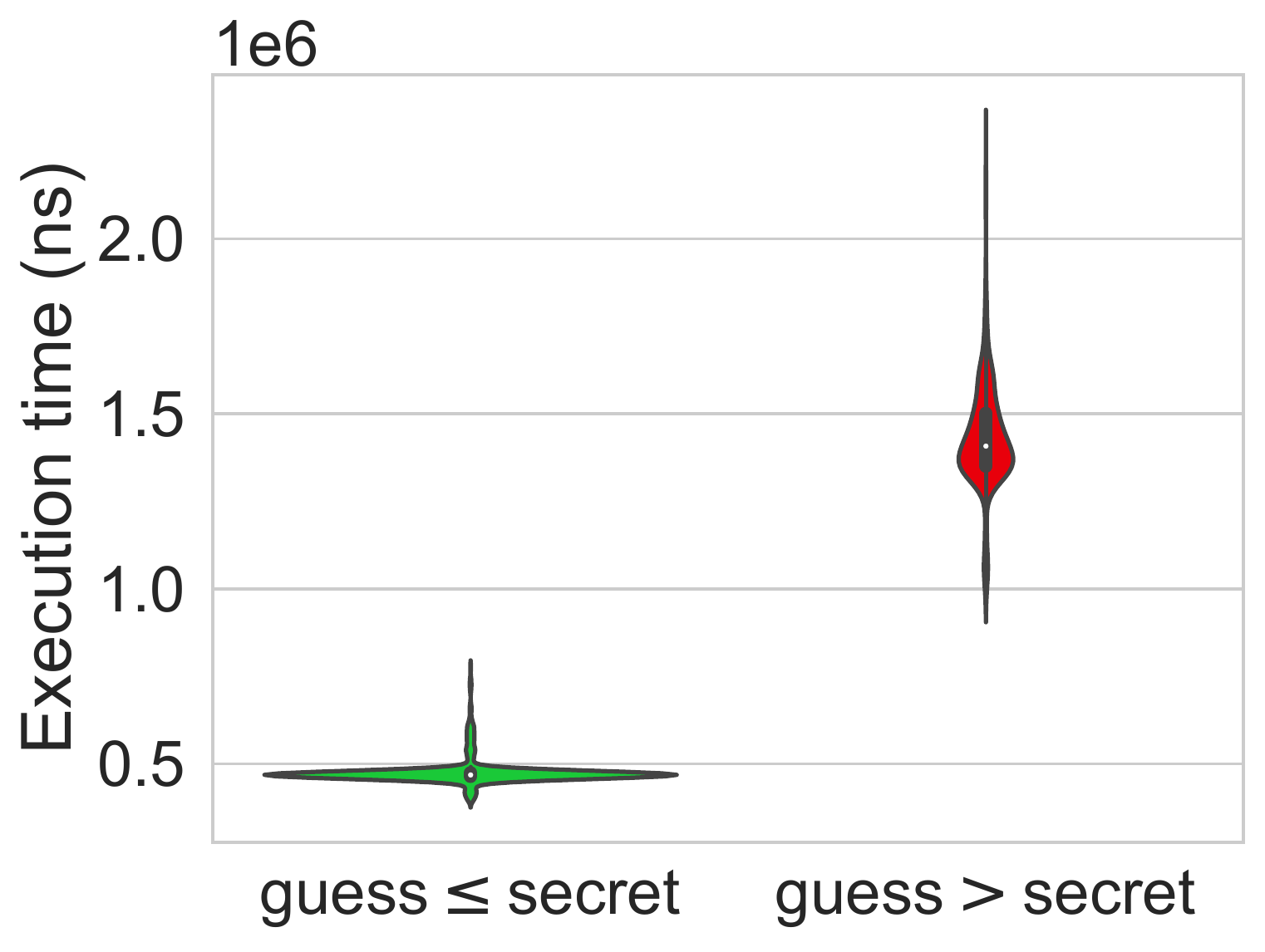}
    \caption{JIT enabled}
     \label{fig:TmethDemo-JITenabled}
  \end{subfigure}
    \begin{subfigure}[b]{0.24\textwidth}
    \centering
    \includegraphics[width=\textwidth]{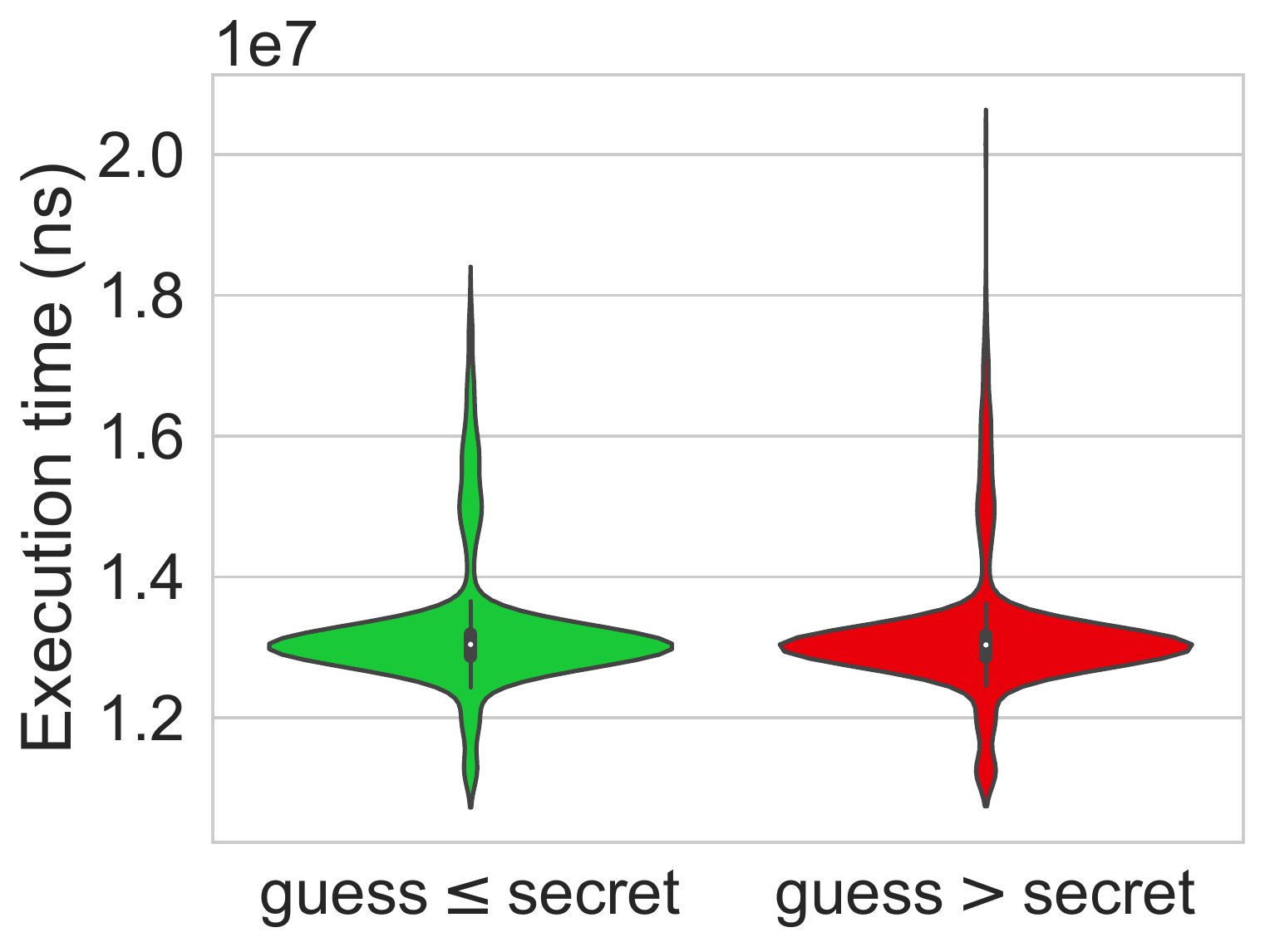}
    \caption{JIT disabled}
     \label{fig:TmethDemo-JITdisabled}
  \end{subfigure}
    \begin{subfigure}[b]{0.24\textwidth}
    \centering
    \includegraphics[width=\textwidth]{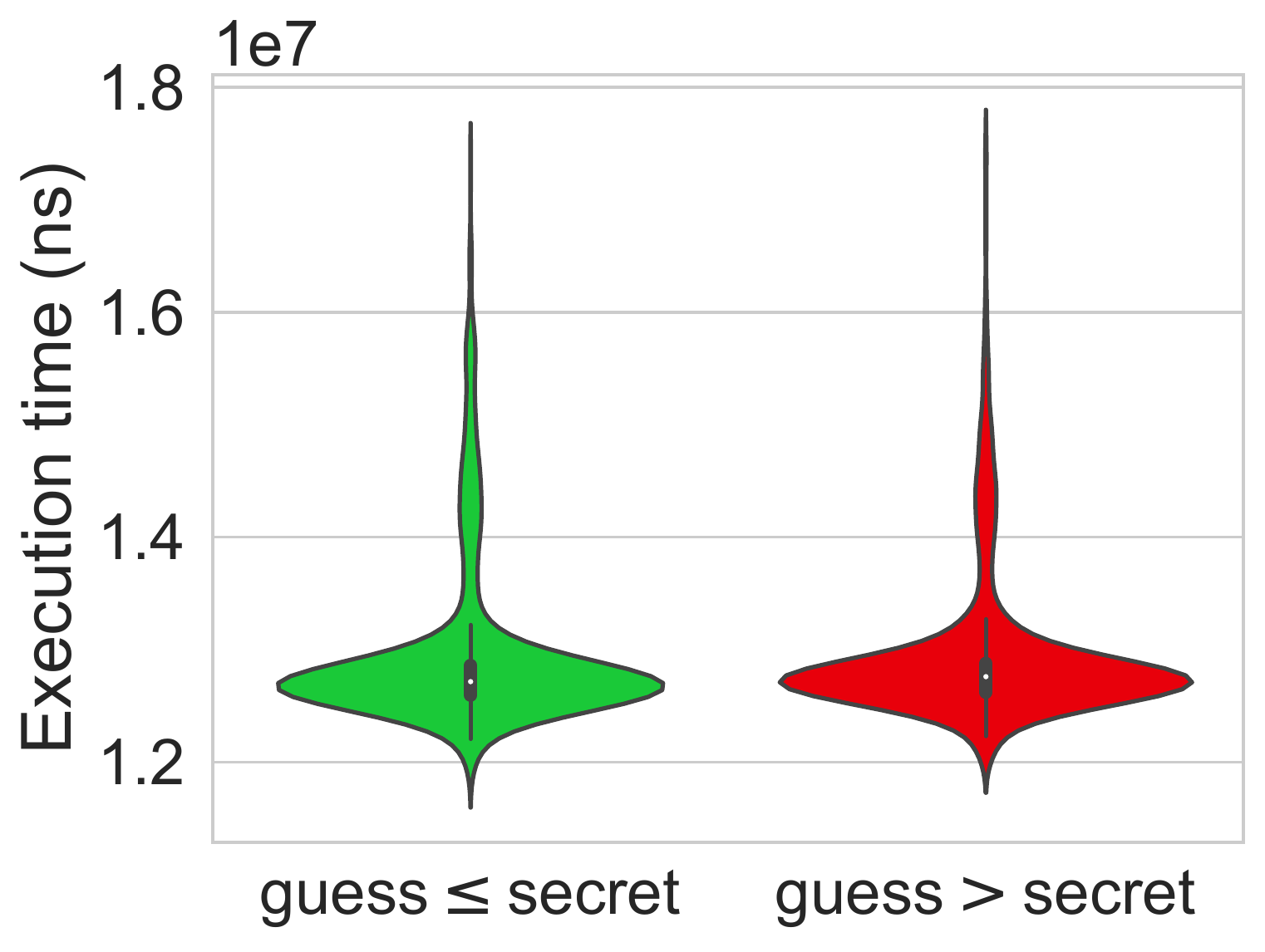}
    \caption{Mitigated}
    \label{fig:mitigationDemocheckSecret}
  \end{subfigure}
  \caption{The \textcolor{blue}{checkSecret} method and its execution time under \textsc{Tbran}}
  \centering
  \label{fig:TmethDemo}
\end{figure*}

\subsection{Full Explanation of the Semantics}

Instruction $\push\ v$, pushes the value $v$ on top of the operand stack.
Instruction $\pop$, just pops the top of the operand stack.
Instruction $\binop\ op$ pops the top two operands from the operand stack and pushes the result of the binary operation $op$ using these operands.
Instruction $\ifeq\ j$ (resp.\ $\ifneq\ j$) pops the top $v$ of the operand stack and
transfers of control to the program point $j$ if $v=0$ (resp. $v\neq 0$), otherwise to the next instruction, i.e., the program point $j+1$.
Instruction $\swap$, swaps the top two values of the operand stack.
Instruction $\store\ x$ (resp.\ $\putt\ y$) pops the top of the operand stack and stores
it in the local  variable $x$ (resp. global variable $y$).
Instruction $\load\ x$ (resp.\ $\get\ y$) pushes the value of the local  variable $x$ (resp.\ global variable $y$), on top of the operand stack.
Instruction $\goto\ j$ unconditionally jumps to program point $j$.
%

Instruction $\return$ ends the execution of the current method,
returns the top value $v$ of the current operand stack,
either by pushing it on top of the operand stack of the caller
and re-executes the caller from the return site if the current method
is not the entry point, or enters a final configuration $(\heap,v)$ if the current method
is  the entry point.

Instruction $\deopt\ \md$ deoptimizes and rolls back to the bytecode in the interpreted mode.
This instruction is only used in native code and inserted by JIT compilers.
Our semantics does not directly model a deoptimization implementation.
Instead, we assume there is a deoptimization oracle $\mathcal{O}$
which takes the current configuration and the meta data $\md$ as inputs,
and reconstructs the configuration (i.e.,  heap $\heap'$, state $s$ and the call stack 
$\cs'$). Furthermore, the bytecode version $\base(m)$ of the method $m$
is restored into the code heap $\cheap$. We assume that the oracle $\mathcal{O}$ results
in the same heap $\heap'$, state $s$ and call stack $\cs'\cdot\cs$
as if the method $m$ were not JIT compiled.

The semantics of method invocation $\invoke\ m'$ depends on the directive $\dir$.
If $\dir$ is $\dir_\emptyset$ 
then the  instructions of $m'$ in the code heap $\cheap$ remain the same.
%
If  $\dir$ is valid, namely,
the optimized version $\mathcal{V}_{m''}$ after applying $\dir$ has larger version number than that of the current version $\mathcal{V}_{m'}$, the new optimized version $m''=\dir(m')$
is stored in the code heap $\cheap$.
After that, it pops the top $|\argv(m')|$
values from the current operand stack, passes them to
the formal arguments $\argv(m')$ of $m'$, pushes
the calling context on top of the call stack
and starts to execute $m'$ in the code heap.

\begin{figure}[t]
\begin{minipage}{0.22\textwidth}
  \centering
\begin{minted}[fontsize=\footnotesize,escapeinside=||,mathescape=true,bgcolor=bg]{java}
  0: |\textcolor{black}{\bf \load} $x_0$|
  1: |\textcolor{black}{\bf \get}|  pin
  2: |\textcolor{black}{\bf \sub}|
  3: |\textcolor{black}{\bf \ifeq} 6|
  4: |\textcolor{blue}{\bf \push} 0|
  5: |\textcolor{blue}{\bf \goto} 8|
  6: |\textcolor{darkgreen}{\bf \push} 1|
  7: |\textcolor{darkgreen}{\bf \goto} 8|
  8: |{\bf \return}|
  \end{minted}
  \caption{Before branch prediction optimization}
  \label{fig:branch-predication-opti-verifyPin1}
\end{minipage}
\quad\begin{minipage}{0.22\textwidth}
    \centering
\begin{minted}[fontsize=\footnotesize,escapeinside=||,mathescape=true,bgcolor=bg]{java}
  0: |\textcolor{black}{\bf \load} $x_0$|
  1: |\textcolor{black}{\bf \get}|  pin
  2: |\textcolor{black}{\bf \sub}|
  3: |\textcolor{black}{\bf \ifeq} 6|
  4: |\textcolor{blue}{\bf \push} 0|
  5: |{\bf \return}|
  6: |\textcolor{darkgreen}{\bf \push} 1|
  7: |\textcolor{darkgreen}{\bf \goto} 5|
  8: |\textcolor{darkgreen}{\bf \goto} 5|
  \end{minted}
  \caption{After branch prediction  optimization}
  \label{fig:branch-predication-opti-verifyPin2}
\end{minipage}
\end{figure}
\subsection{Formal Definition of the New Method $m'$ After Branch Prediction}
If the profiling data show that the program favors the else-branch instead of the if-branch,
the branch prediction optimization
transforms the method $m$ into a new method $m_1$ as follows:
\begin{itemize}
  \item $m_1[i]$ becomes $\ifeq\ j''$ (resp. $\ifneq\ j''$) if $m[i]$ is $\ifeq\ j$ (resp. $\ifneq\ j$),
  where $j''=|m|-j'+j-1$ is the point of the first instruction of the if-branch $B_{\tt t}'$ in $m_1$;
  \item the if-branch $B_{\tt t}'$ in $m_1$ is the if-branch  $B_{\tt t}$, but is moved to the end of the method and appended with  \goto\ $j-1$, namely,
     $m_1[|m|-j'+j-1,|m-2|]=m[j,j'-1]$ and $m_1[|m-1|]= \goto\ j-1$;
  \item the else-branch $B_{\tt f}'$ in $m_1$ is the else-branch  $B_{\tt f}$, where the last instruction \goto\ $j'$ is removed, namely,   $m_1[i+1,j-2]=m[i+1,j-2]$;
  \item furthermore, the target points of other conditional and unconditional jumps are revised accordingly.
\end{itemize}

\begin{figure}[t]
	\centering
	\includegraphics[width=.4\textwidth]{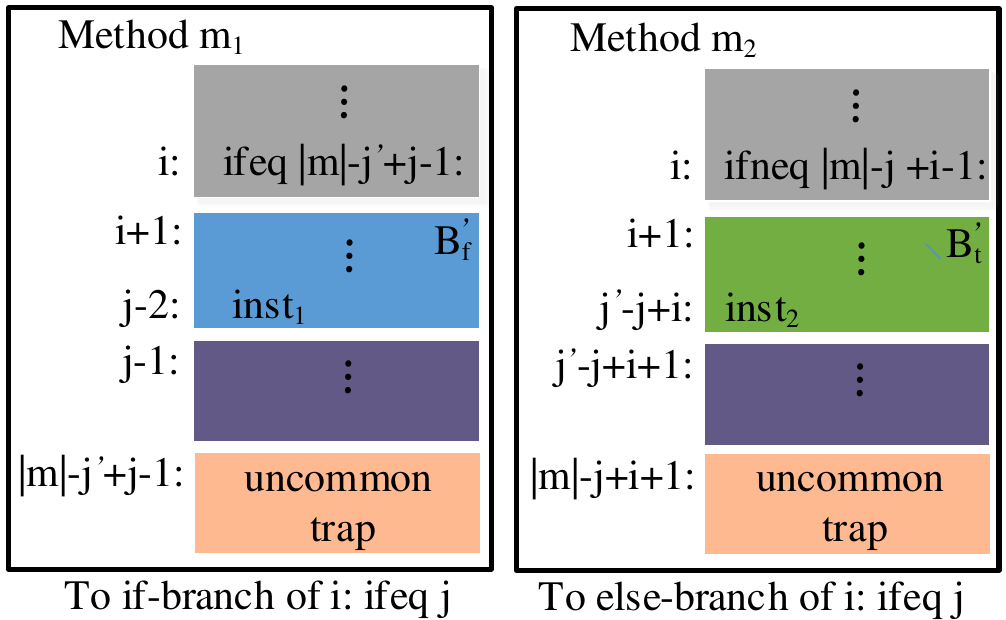}
	\caption{Optimistic compilation optimization}\label{fig:optimistic-compilation-opti}
\end{figure}

\begin{example} \label{example:branch-predication-opti}
Consider the method \textcolor{blue}{verifyPin} shown in Figure~\ref{fig:branch-predication-opti-verifyPin1}, where
the else-branch is at lines 4--5 and the if-branch is at lines 6--7. Note that the instruction at line 7 is added to balance the execution time of the two branches, which is done by manipulating the bytecode.

We can get $\transform_{\tt bp}(\mbox{\textcolor{blue}{verifyPin}},3,\elseb)$, as shown in Figure~\ref{fig:branch-predication-opti-verifyPin2},
where the else-branch is at line 4 and the if-branch is at lines 6--8. Obviously, the optimization unbalances the execution time of the two branches. Note that the instruction at line 8 is dead code, hence may be removed by other optimizations (e.g., peephole optimization),
but the execution time of the two branches is still unbalanced.
\end{example}

\subsection{Methods $m_1$ amd $m_2$ after Optimistic Compilation}
Figure~\ref{fig:optimistic-compilation-opti} (left-part) shows
the method $m$' after the optimistic compilation optimization
when the profiling data show that the if-branch 
almost never gets executed.

Figure~\ref{fig:optimistic-compilation-opti} (right-part) shows
the method $m$' after the optimistic compilation optimization
when the profiling data show that the else-branch 
almost never gets executed.

\subsection{Proof of Theorem~\ref{thm:soundness}}
{\bf Theorem~\ref{thm:soundness}}.
Given a program $P$, if $P$ is constant-time and $(\prot_1,\prot_2,\sig_P)\rhd P$,
then $P$ is JIT-constant-time under $(\prot_1,\prot_2)$-schedules.

\begin{proof}
Consider the program $P$. Assume $P$ is constant-time and $(\prot_1,\prot_2,\sig_P)\rhd P$
holds. To prove that $P$ is JIT-constant-time under $(\prot_1,\prot_2)$-schedules, we first introduce some notations.

Recall that for every method $m$ and every valid directive $\dir\in\Dir_m$, we denoted
by $\dir(m)$ the new optimized version of $m$ after applying the directive $\dir$.
For a sequence of valid directives $\dir_1\cdots\dir_n$ of $m$,
we denote by $[\dir_1\cdots\dir_n](m)$ the latest optimized version of $m$ after sequentially
applying the directives $\dir_1\cdots\dir_n$, where $[\epsilon](m)=m$.

For each method $m$ with $\sig_P(m)=(\pt_1,\heapt_1,\lt_1)\hookrightarrow_{m}(\heapt_1', \tau)$,
two configurations $c_0=(\cheap,\heap,\langle 0,m, \rho, \epsilon\rangle,\epsilon)$
and $c_0'=(\cheap,\heap',\langle 0,m, \rho', \epsilon\rangle,\epsilon)$
are $\sig_P(m)$-equivalent, denoted by $c_0\simeq_{\sig_P(m)}c_0'$,
if $\heap$ and $\heap'$ agree on the variables that have high security
level $\high$ in $\heapt_1$, and $\rho$ and $\rho'$ agree on the variables that have high security $\high$
level in $\lt_1$.

To prove that $P$ is JIT-constant-time under $(\prot_1,\prot_2)$-schedules, we first prove the following 
lemma: 

\begin{lemma}\label{lem:soundness1step}
Consider a method $m$ and two configurations $c_0=(\cheap,\heap,\langle 0,m, \rho, \epsilon\rangle,\epsilon)$
and $c_0'=(\cheap,\heap',\langle 0,m, \rho', \epsilon\rangle,\epsilon)$  such that
$c_0\simeq_{\sig_P(m)}c_0'$.
Given a sequence of valid directives $\dir_1\cdots\dir_n$ of $m$ w.r.t. the policy
$(\prot_1,\prot_2)$, let $c_2$ and $c_2'$ be two configurations such that
 $c_2=(\cheap,\heap,\langle 0, [\dir_1\cdots\dir_n](m), \rho, \epsilon\rangle,\epsilon)$
and $c_2'=(\cheap,\heap',\langle 0,[\dir_1\cdots\dir_n](m), \rho', \epsilon\rangle,\epsilon)$.
We have:

if $\cost(c_0 \Downarrow_{\dir^{\star}_{\emptyset}} c_1)=\cost(c_0' \Downarrow_{\dir^{\star}_{\emptyset}} c_1')$,

then $\cost(c_2 \Downarrow_{\dir^{\star}_{\emptyset}} c_1)=\cost(c_2' \Downarrow_{\dir^{\star}_{\emptyset}} c_1')$.
\end{lemma}

We prove this lemma by induction on the length of the sequence $\dir_1\cdots\dir_n$.
The base case follows from the fact that $[\epsilon](m)=m$.
We consider the inductive step $n\geq 1$. 

Let $c_3=(\cheap,\heap,\langle 0, [\dir_1\cdots\dir_{n-1}](m), \rho, \epsilon\rangle,\epsilon)$
and $c_3'=(\cheap,\heap',\\\langle 0,[\dir_1\cdots\dir_{n-1}](m), \rho', \epsilon\rangle,\epsilon)$.
Then, $c_3\simeq_{\sig_P(m)}c_3'$.
By applying the induction hypothesis: we get that $\cost(c_3 \Downarrow_{\dir^{\star}_{\emptyset}} c_1)=\cost(c_3' \Downarrow_{\dir^{\star}_{\emptyset}} c_1')$.
If $\dir_n=\dir_{\emptyset}$,  the result immediately follows. We consider $\dir_n=(t,\omega)$.

Let $c_4=(\cheap,\heap,\langle 0, t([\dir_1\cdots\dir_{n-1}](m)), \rho, \epsilon\rangle,\epsilon)$
and $c_4'=(\cheap,\heap',\langle 0,t([\dir_1\cdots\dir_{n-1}](m)), \rho', \epsilon\rangle,\epsilon)$.
Then, $c_4\simeq_{\sig_P(m)}c_4'$.

Since the directive $\dir_n$ respects the $(\prot_1,\prot_2)$ policy and $(\prot_1,\prot_2,\sig_P)\rhd P$,
all the methods specified in $t$ can be invoked only with the path context $\low$. 
This implies that the subsequences of executed instructions of the inlined methods are the same in the JIT-executions
$c_4 \Downarrow_{\dir^{\star}_{\emptyset}} c_1$ and $c_4' \Downarrow_{\dir^{\star}_{\emptyset}} c_1'$,
we get that $\cost(c_4 \Downarrow_{\dir^{\star}_{\emptyset}} c_1)=\cost(c_4' \Downarrow_{\dir^{\star}_{\emptyset}} c_1')$.
Note that we have already assumed that the cost equivalence of bytecode instructions are preserved in native code.

Let us now consider the sequence $\omega$ of optimizations to branch points. 
As $(\prot_1,\prot_2,\sig_P)\rhd P$, no branch points whose the path context or condition is $\high$
can appear in $\omega$, thus  all the branches of the branch points of $\omega$
are the same in $t([\dir_1\cdots\dir_{n-1}](m))$ and $[\dir_1\cdots\dir_{n}](m)$ (module the code format, i.e., bytecode vs. native code if $n=1$ and $\cheap(m)$ is bytecode).
Since the subsequences of executed branches of the branch points with the path context $\low$
are the same in the JIT-executions
$c_2 \Downarrow_{\dir^{\star}_{\emptyset}} c_1$ and $c_2' \Downarrow_{\dir^{\star}_{\emptyset}} c_1'$,
 we get that $\cost(c_2 \Downarrow_{\dir^{\star}_{\emptyset}} c_1)=\cost(c_2' \Downarrow_{\dir^{\star}_{\emptyset}} c_1')$.

Now, we start to prove that $P$ is JIT-constant-time under $(\prot_1,\prot_2)$-schedules.

Consider a pair of initial configurations $(c_0,c_0')$ of $P$
with $c_0\simeq_{\tt pub} c_0'$ and $c_0\simeq_{\tt ch} c_0'$.
Let $c_1$ and $c_1'$ be the configurations obtained from $c_0$ and $c_0'$
by replacing the code heap with the bytecode version.
Then, there exists a valid $(\prot_1,\prot_2)$-schedule $\dir^\star$ such that
the code heap in $c_0$ and $c_0'$ is equal to the bytecode version after applying $\dir^\star$.
Since $P$ is constant-time, we get that $\cost(c_1 \Downarrow_{\dir^{\star}_{\emptyset}} c)=\cost(c_1' \Downarrow_{\dir^{\star}_{\emptyset}} c').$
By Lemma~\ref{lem:soundness1step}, we have:
$\cost(c_0 \Downarrow_{\dir^{\star}_\emptyset} c)=\cost(c_0' \Downarrow_{\dir^{\star}_\emptyset} c')$.

Consider a pair of valid $(\prot_1,\prot_2)$-schedules $\dir^{\star}_1$ and $\dir^{\star}_2$ for $c_0$ and $c_0'$.
Then, only methods $m\in\Methods\setminus\prot_1$ can be JIT compiled or inlined,
namely, only methods that are never invoked with the path context $\high$ can be JIT compiled or inlined.
Since $(\prot_1,\prot_2,\sig_P)\rhd P$ and $c_0\simeq_{\tt pub} c_0'$,
we obtain that $\dir^{\star}_1$ and $\dir^{\star}_2$ must be the same.

We show that $\cost(c_0 \Downarrow_{\dir^{\star}_1} c)=\cost(c_0' \Downarrow_{\dir^{\star}_1} c')$
by induction on the number of non-$\dir^{\star}_\emptyset$ directives in $\dir^{\star}_1$.
The base case follows from the fact that $\cost(c_0 \Downarrow_{\dir^{\star}_\emptyset} c)=\cost(c_0' \Downarrow_{\dir^{\star}_\emptyset} c')$.
For the inductive step, we assume that ${\dir^{\star}_3}$
is the $(\prot_1,\prot_2)$-schedule obtained from ${\dir^{\star}_1}$
by replacing the last directive $\dir\in\Dir_m$ of ${\dir^{\star}_1}$ with the directive $\dir_\emptyset$. 
By applying the induction hypothesis, we get that 
$\cost(c_0 \Downarrow_{\dir^{\star}_3} c)=\cost(c_0' \Downarrow_{\dir^{\star}_3} c')$.
By Lemma~\ref{lem:soundness1step}, we get that
$\cost(c_0 \Downarrow_{\dir^{\star}_1} c)=\cost(c_0' \Downarrow_{\dir^{\star}_1} c')$.
 \end{proof}

\end{document}